\documentclass[aip,amsmath,amssymb,preprint,longbibliography]{revtex4-1}

\usepackage{graphicx}
\graphicspath{{figs_pof/}{../../paper_ijhmt/}{../analysis/}}
\usepackage{booktabs}
\usepackage{multirow}
\usepackage{bm}
\usepackage{subcaption}
\usepackage{tabularx}
\usepackage{placeins}
\usepackage{float}
\usepackage{needspace}
\usepackage{comment}

\newcommand{\hide}[1]{}
\newcommand{\RAF}{\mathrm{RAF}}
\newcommand{\Dafr}{Da_{fr}}

\begin{document}

\title{Finite-rate chemistry in a hypersonic wall model:
heat-flux selection, non-monotone rate response and transport-closure effects}

\author{Jingchao Zhang}
\affiliation{Hangzhou International Innovation Institute, Beihang University,
Hangzhou 311115, China}
\affiliation{School of Aeronautic Science and Engineering, Beihang University,
Beijing 100191, China}
\email{jingchaozhang@buaa.edu.cn}

\date{\today}

\begin{abstract}
This study examines how the chemical-reaction timescale alters wall heat flux and shear stress predicted by a one-dimensional wall model with finite-rate chemistry. Calculations for high-enthalpy air span $3 \le Ma_e \le 25$, $2 \le p_e \le 100~\mathrm{kPa}$, and a broad range of edge and wall temperatures. All forward and reverse reaction rates are multiplied by a common factor $\Gamma$, which changes the chemical timescale while preserving the equilibrium constants and thereby connects the frozen and infinitely fast reaction-rate limits within the same governing equations. Taking a 1\% difference in wall heat flux between the frozen and finite-rate solutions as the threshold for retaining finite-rate chemistry, two quantities obtained from the frozen solution provide practical selection criteria: $Da_{\mathrm{fr}}=t_{\mathrm{shear}}/t_{\mathrm{ch}} \ge 1.93\times10^{3}$ and $T_{\max}\ge3760~\mathrm{K}$. The wall heat flux does not generally vary monotonically with reaction rate; instead, it can decrease below both the frozen and fast-chemistry predictions before recovering at higher rates. At $Ma_e\ge20$, the median wall-heat-flux errors of the frozen and high-rate approximations are 7.1\% and 27.6\%, respectively. Chemistry also produces substantial changes in wall heat flux and shear stress, whereas the Reynolds analogy factor responds much less because the momentum- and enthalpy-side wall-law shifts nearly cancel. This cancellation is sensitive to the turbulent transport closure. These results show that wall heat flux at finite reaction rates cannot in general be inferred from the frozen and infinitely fast reaction-rate limits and provide
frozen-state indicators for determining when finite-rate chemistry should be retained
in hypersonic wall modelling.
\end{abstract}

\keywords{hypersonic boundary layer; finite-rate chemistry; wall model;
compressible wall turbulence; turbulent transport}

\maketitle

\section{Introduction}
\label{sec:intro}
Hypersonic flows can generate very high gas temperatures through shock compression
and viscous dissipation, activating molecular dissociation, recombination, and other
high-temperature chemical processes \citep{candler2019rate}. Because these reactions
proceed at finite rates, the thermochemical state of the gas need not remain in chemical
equilibrium as it is transported through the boundary layer
\citep{candler2019rate,passiatore2021finite,passiatore2022thermochemical}. The
competition between chemical-reaction and near-wall transport timescales can consequently
modify the temperature, composition, and enthalpy distributions, as well as wall heat flux,
wall shear stress, and the conventional scaling of the near-wall velocity and thermal fields
\citep{di2021direct,liu2022high,li2022wall}. Resolving these coupled processes throughout
the near-wall region is computationally expensive in high-Reynolds-number simulations.
Wall modelling therefore replaces the fully resolved inner layer with a reduced description
that relates the outer flow state to the wall fluxes while retaining the physical effects
required for the target conditions \citep{kawai2012wallmodeling}. In high-enthalpy
applications, this requires a choice among frozen, finite-rate nonequilibrium, and
equilibrium thermochemical treatments, which differ in both physical assumptions and
computational cost \citep{candler2019rate,muto2021wall,huang2026wallmodelled}. A central
question is therefore how the chemical-reaction timescale affects the wall quantities and
near-wall scalings predicted by the model, and under what conditions finite-rate chemistry
must be retained rather than replaced by a limiting thermochemical description.

Direct numerical simulations (DNS) have established that finite-rate chemistry and
multicomponent transport can remain dynamically important in hypersonic boundary layers
at flight enthalpy, allowing the gas to remain thermochemically out of equilibrium over a
substantial part of the boundary layer
\citep{di2021direct,passiatore2021finite,passiatore2022thermochemical}. Subsequent studies
have clarified several consequences of this nonequilibrium, including its influence on
boundary-layer instability
\citep{marxen2014direct,chen2021secondary}, similarity of chemical behaviour across
reacting layers \citep{passiatore2025chemistry}, and turbulence--chemistry interactions
identified through direct and probability-density-function analyses
\citep{williams2025turbulence,wang2026probability}. Finite-rate chemistry also modifies
the mean near-wall structure: changes in thermochemical state affect the velocity and
thermal fields \citep{di2021direct,liu2022high}, while the resulting influence on wall
heat transfer has been quantified in high-enthalpy turbulent boundary layers
\citep{li2022wall,liu2023walltemperature,fratini2026walltemperature}. These studies
establish that thermochemical nonequilibrium can affect near-wall scaling, turbulent
transport, and wall heat transfer, and that the magnitude of these effects depends on the
flow and wall thermal conditions. Most existing studies, however, examine a prescribed
chemistry model at selected thermodynamic conditions. A systematic relation between the
chemical-reaction timescale and the resulting wall response, together with a quantitative
criterion for frozen-chemistry adequacy in reduced wall modelling, therefore remains to be
established.

To represent these near-wall effects at a cost compatible with wall-modelled LES, reduced
wall models have incorporated progressively more of the physics relevant to compressible
and high-enthalpy flows. Mass injection has been described through mixing-length and
asymptotic momentum formulations, which primarily modify the logarithmic velocity-law
intercept \citep{stevenson1963law,freire1995velocity}. Compressibility and strong wall
cooling have been treated through total-shear-stress velocity transformations, algebraic
eddy-viscosity corrections, and corresponding transformations of the temperature field
\citep{griffin2021velocity,chen2024improved,xu2026temperature}. Generalised recovery
factors and temperature--velocity relations have been developed to account for strong
compressibility and wall-thermal effects
\citep{zhang2014generalized,mo2024new}, while reacting high-enthalpy extensions have
introduced enthalpy--velocity formulations that account explicitly for chemical energy
storage and dissociation \citep{li2025improved}. Other formulations determine the wall
fluxes by matching the inner solution to the resolved flow \citep{xu2025flux} or by
solving coupled momentum and energy ordinary differential equations
\citep{huang2025ordinary}. Reacting-flow wall models have also compared frozen and
equilibrium thermochemical treatments \citep{muto2021wall}, and recent high-enthalpy
wall-modelled LES formulations have incorporated catalytic wall chemistry
\citep{huang2026wallmodelled}. These developments substantially extend wall modelling
beyond calorically perfect-gas assumptions and allow high-enthalpy thermochemical effects
to be represented at reduced cost. However, the thermochemical model is generally
prescribed before the wall solution is obtained. Consequently, existing formulations do
not provide a quantitative basis for deciding when frozen chemistry is adequate, when
finite-rate chemistry must be retained, or how that decision changes with the turbulent
heat- and species-transport closure.

The present study addresses this model-selection problem by varying the chemical-reaction
timescale within a common finite-rate wall-model formulation. All forward and reverse
reaction-rate coefficients are multiplied by a uniform factor $\Gamma$, so that the
chemical timescale is changed while the equilibrium constants, and hence the equilibrium
composition at a given thermodynamic state, remain unchanged. The limits
$\Gamma\to0$ and $\Gamma\to\infty$ correspond, respectively, to frozen chemistry and
the infinitely fast reaction-rate limit of the same governing equations, while intermediate
values provide a continuous variation of reaction timescale under fixed imposed flow
conditions. This construction separates finite-rate effects from changes in equilibrium
thermodynamics and allows the wall response to be examined consistently between the two
limiting regimes. The analysis focuses on three questions: whether quantities available
from the frozen solution can identify conditions for which finite-rate chemistry must be
retained to predict wall heat flux within a prescribed tolerance; how wall heat flux and
related near-wall quantities vary as the reaction timescale is changed; and how the
resulting chemistry sensitivity and Reynolds analogy factor depend on the turbulent heat-
and species-transport closures.

\section{Methodology}
\label{sec:method}

\subsection{Wall model and thermochemical formulation}
\label{sec:wallmodel}

The analysis uses a one-dimensional wall-layer model that relates a prescribed
matching-height state to the corresponding wall momentum and heat fluxes~\citep{10.1063/5.0344351}. The model assumes a
steady, locally parallel boundary layer with uniform pressure across the wall-normal
direction, $p(y)=p_e$. Streamwise derivatives and the edge-pressure-gradient term are
neglected. The present study considers an impermeable, non-ablating wall, so that the
wall-normal mass flux is zero, $M_w=\rho_wv_w=0$. Continuity then gives $\rho v=0$ throughout
the modelled layer, and the momentum, energy and species equations reduce to
\begin{subequations}
\label{eq:wallmodel}
\begin{align}
0={}&\frac{\mathrm d}{\mathrm dy}
\left(\mu_{\mathrm{eff}}\frac{\mathrm du}{\mathrm dy}\right),
\label{eq:wallmodel_mom}\\
0={}&\frac{\mathrm d}{\mathrm dy}
\left(\lambda_{\mathrm{eff}}\frac{\mathrm dT}{\mathrm dy}\right)
-\sum_{s=1}^{N_s}c_{p,s}J_s\frac{\mathrm dT}{\mathrm dy}
+\mu_{\mathrm{eff}}\left(\frac{\mathrm du}{\mathrm dy}\right)^2
-\sum_{s=1}^{N_s}h_s\dot{\omega}_s
+\frac{\mathrm d}{\mathrm dy}\left[
\left(\mu+\frac{\mu_t}{\sigma_K}\right)\frac{\mathrm dK}{\mathrm dy}\right],
\label{eq:wallmodel_energy}\\
0={}&-\frac{\mathrm dJ_s}{\mathrm dy}+\dot{\omega}_s,
\qquad s=1,\ldots,N_s-1 .
\label{eq:wallmodel_species}
\end{align}
\end{subequations}
Here $u$ and $v$ are the streamwise and wall-normal velocities, $\rho$ is the mixture density,
$\mu$ and $\lambda$ are the molecular viscosity and thermal conductivity, and $\mu_t$ is the
eddy viscosity. The subscript $e$ denotes a quantity at the matching height, $N_s$ is the
number of species, and the edge Mach number is $Ma_e=u_e/a_e$ with $a_e$ the edge sound speed.
The effective momentum and thermal transport coefficients are
\[
\mu_{\mathrm{eff}}=\mu+\mu_t,
\qquad
\lambda_{\mathrm{eff}}=\lambda+\frac{\mu_tc_p}{Pr_t},
\]
where $Pr_t$ is the turbulent Prandtl number. Species transport is represented by the
corrected mixture-averaged flux
\begin{equation}
\widetilde J_s=-\left(\rho D_s+\frac{\mu_t}{Sc_t}\right)
\frac{\mathrm dY_s}{\mathrm dy},\qquad
J_c=-\sum_{r=1}^{N_s}\widetilde J_r,\qquad
J_s=\widetilde J_s+Y_sJ_c,
\label{eq:species_flux}
\end{equation}
where $c_p$ and $c_{p,s}$ are the mixture and species heat capacities, $h_s$ is the species
enthalpy, $\dot{\omega}_s$ is the net mass-production rate, $D_s$ is the mixture-averaged
molecular diffusivity, $Sc_t$ is the turbulent Schmidt number, and the correction flux $J_c$
enforces $\sum_sJ_s=0$. The two species terms in \eqref{eq:wallmodel_energy} are equivalent to
$-\mathrm d(\sum_sh_sJ_s)/\mathrm dy$ after use of \eqref{eq:wallmodel_species}, so that
molecular and turbulent species transport enter the wall-layer enthalpy balance explicitly.

The baseline turbulent-transport closure uses $Pr_t=Sc_t=0.89$. The eddy viscosity is
$\mu_t=\rho l_m^2|\mathrm du/\mathrm dy|$, with the mixing length given by the
compressibility-corrected closure
\begin{equation}
l_m=\min\left\{\kappa y\left[1-\exp\left(
-\frac{y^*}{26+C_{ML}Ma_\tau}\right)\right],\ 0.09\delta_{99}\right\},
\qquad
y^*=y^+\frac{\mu_w}{\mu}\sqrt{\frac{\rho}{\rho_w}},
\label{eq:mixing_length}
\end{equation}
where $\tau_w$ is the wall shear stress, $u_\tau=\sqrt{|\tau_w|/\rho_w}$ is the friction
velocity, $y^+=\rho_wu_\tau y/\mu_w$, $a_w$ is the wall sound speed, $Ma_\tau=u_\tau/a_w$, and
the subscript $w$ denotes a wall value. The model constants are $\kappa=0.41$ and $C_{ML}=37$
\citep{hasan2023incorporating,trettel2016mean}. The algebraic turbulent-kinetic-energy
contribution appearing in \eqref{eq:wallmodel_energy} is evaluated from the local eddy
viscosity and velocity gradient as $K=\mu_t|\mathrm du/\mathrm dy|/(\rho\sqrt{C_\mu})$, with
$\sqrt{C_\mu}=0.3$ and $\sigma_K=1$ \citep{bradshaw1967calculation}; no separate transport
equation for turbulent kinetic energy is solved.

The wall and matching-height boundary conditions are
\begin{equation}
\begin{aligned}
u&=0,\quad T=T_w,\quad J_s=0 && (y=0),\\
u&=u_e,\quad T=T_e,\quad Y_s=Y_{s,e} && (y=y_m).
\end{aligned}
\label{eq:wallmodel_bc}
\end{equation}
The imposed edge composition is air, with $Y_{\mathrm{N_2},e}=0.79$ and
$Y_{\mathrm{O_2},e}=0.21$ and all other edge mass fractions equal to zero. This composition is
held fixed when the chemical timescale is varied. The wall shear stress and wall heat flux are
\begin{equation}
\tau_w=\left.\mu\frac{\mathrm du}{\mathrm dy}\right|_w,
\qquad
q_w=\left.\left(\lambda\frac{\mathrm dT}{\mathrm dy}
-\sum_{s=1}^{N_s}h_sJ_s\right)\right|_w,
\label{eq:wall_fluxes}
\end{equation}
where $q_w>0$ denotes heat transfer from the gas to the wall. The corresponding skin-friction
coefficient, Stanton number and enthalpy driving potential are defined by
\begin{equation}
C_f=\frac{2\tau_w}{\rho_eu_e^2},\qquad
St=\frac{q_w}{\rho_eu_e\Delta h},\qquad
\Delta h=h_{aw}-h_w,
\label{eq:transfer_coefficients}
\end{equation}
where $Pr_e=\mu_ec_{p,e}/\lambda_e$, $h_{aw}=h_e+Pr_e^{1/3}u_e^2/2$ is the nominal
recovery-enthalpy estimate used in the Stanton-number normalisation, and $h_w$ is the
wall-mixture enthalpy. For a stated pairing from state $a$ to state $b$,
$\delta\ln X=\ln(X_b/X_a)$; when this increment is reported in per cent, the displayed
quantity is $100\,\delta\ln X$. The matching height is $y_m=1.4\widehat\delta$, where
$\widehat\delta$ is the reference-enthalpy compressible flat-plate estimate at
$x_{\mathrm{ref}}=1$~m, implemented as in Zhang et al.~\citep{ZHANG2025109915} from the
correlation of Eckert~\citep{eckert1955engineering}. The boundary-layer thickness
$\delta_{99}$ in the outer cap of \eqref{eq:mixing_length} is updated during iteration from
the location at which $u=0.99u_e$.

The gas is treated as a thermally perfect mixture, $p=\rho R_mT$ with
$R_m=R_u\sum_sY_s/M_s$, where $R_u$ is the universal gas constant and $M_s$ is the molar mass
of species $s$; the mixture heat capacity and enthalpy are $c_p=\sum_sY_sc_{p,s}$ and
$h=\sum_sY_sh_s$. Species heat capacities and enthalpies are evaluated from the piecewise
NASA-9 polynomial representation \citep{mcbride2002nasa}. Pure-species viscosities and
conductivities follow kinetic theory using the Stockmayer potential and the modified Eucken
relation, with molecular parameters and collision integrals from the CHEMKIN transport
database \citep{kee1986fortran}; mixture transport coefficients use the Wilke rule
\citep{wilke1950viscosity}, binary diffusion follows the Chapman--Enskog relation, and the
mixture-averaged diffusivity uses the Hirschfelder--Curtiss approximation
\citep{hirschfelder1964molecular}. The wall-layer equations are solved for the primary
variables $\bm q=[u,T,Y_1,\ldots,Y_{N_s-1}]^{\mathrm T}$, with the final mass fraction
obtained from $\sum_sY_s=1$, on a geometrically stretched wall-normal grid as a fully coupled
nonlinear boundary-value problem using a block-tridiagonal Newton method with
pseudo-transient continuation. Numerical tolerances, grid convergence and continuation
details are reported in the supplementary material and by Zhang
et al.~\citep{ZHANG2025109915}.

Chemical source terms are evaluated using a five-species finite-rate air
chemistry model comprising N$_2$, O$_2$, NO, N and O, with reaction-rate
coefficients following Park~\citep{park1989assessment}, including air dissociation, recombination and
exchange reactions.

\subsection{Controlled variation of the chemical timescale}
\label{sec:gamma}

To isolate the influence of the chemical-reaction timescale from changes in equilibrium
thermodynamics, all forward and reverse reaction-rate coefficients are multiplied by the same
factor $\Gamma$,
\begin{equation}
k_f\to\Gamma k_f,\qquad k_b\to\Gamma k_b .
\label{eq:gamma}
\end{equation}
Because the same multiplier is applied to both directions of every reaction, the equilibrium
constants $K_{eq}=k_f/k_b$ remain unchanged. The equilibrium composition at a prescribed
thermodynamic state is therefore unaffected, whereas the characteristic time required for the
chemical system to approach that equilibrium changes in proportion to the reaction rates. In
this way, $\Gamma$ acts as a controlled parameter for the chemical timescale without altering
the underlying equilibrium thermodynamics.

The two limiting values have direct physical interpretations. As $\Gamma\to0$, the chemical
source terms become negligible on the wall-layer transport timescale and the solution
approaches the frozen-chemistry limit. As $\Gamma\to\infty$, chemical relaxation becomes
increasingly rapid and the solution approaches the infinitely fast reaction-rate limit of the
same finite-rate governing equations. This latter state is not imposed through a separate
equilibrium-composition model; it is obtained as the high-rate limit of the finite-rate
boundary-value problem. Intermediate values of $\Gamma$ therefore trace how the wall-layer solution changes
as the chemical timescale varies between these two limits.

Each primary thermodynamic condition is evaluated at three chemistry levels: frozen chemistry,
the reference finite-rate mechanism at $\Gamma=1$, and an accelerated finite-rate state at
$\Gamma=100$. A subset of $778$ conditions additionally includes a converged high-rate state at
$\Gamma=10^7$, which is used to approximate the $\Gamma\to\infty$ limit. The $\Gamma=1$ solution retains the original kinetic rates of the
adopted reaction mechanism, whereas the larger values are used only to probe the response to
progressively shorter chemical timescales. The resulting set of solutions allows the
finite-rate wall response to be compared directly with both limiting descriptions while
keeping the imposed flow state, reaction pathways and equilibrium constants fixed. The common rate multiplier changes the overall chemical timescale but does not
represent uncertainty in individual kinetic parameters or relative reaction pathways; a numerical verification
that every active-species chemical time scales as $\Gamma^{-1}$ is given in the supplementary
material.

\subsection{Frozen-state timescale diagnostic}
\label{sec:da}

To relate the wall response to the competition between chemical and near-wall transport
timescales, a Damk\"ohler-type diagnostic is constructed from the frozen solution. The local
chemical timescale is defined from the species source-term Jacobian,
\[
\mathbf J=\rho^{-1}\left(\frac{\partial\bm{\dot\omega}}{\partial\mathbf Y}\right)_{T,p},
\]
where $\bm{\dot\omega}$ is the vector of species mass-production rates and $\mathbf Y$ is the
species mass-fraction vector. For each active species, a characteristic chemical time is
estimated from the corresponding diagonal Jacobian entry. The chemical timescale used here is
the longest of these active-species times,
\begin{equation}
t_{ch}=\max\left\{\,1/|J_{ss}|\;:\;Y_s>10^{-4}\right\}.
\label{eq:tch}
\end{equation}
The corresponding wall-layer transport timescale is taken as the local shear time,
\begin{equation}
t_{shear}=\frac{1}{|\mathrm du/\mathrm dy|}.
\label{eq:tshear}
\end{equation}
Both quantities are evaluated at $y^+=100$ in the frozen solution, and their ratio is denoted
\begin{equation}
\Dafr=\frac{t_{shear}}{t_{ch}}.
\label{eq:dafr}
\end{equation}
Larger values of $\Dafr$ therefore correspond to chemical relaxation that is faster relative
to the local shear timescale, whereas smaller values indicate a greater separation towards the
frozen-chemistry limit.

The use of the frozen solution is deliberate: $\Dafr$ is intended as a diagnostic available
before a finite-rate wall-model calculation is performed. The longest active-species timescale
represents the slow end of the local chemical response, while the mass-fraction cutoff
excludes trace species whose Jacobian times are not representative of the dynamically relevant
mixture.

This definition should be interpreted as a local organising parameter rather than as a unique
chemical relaxation timescale. The diagonal quantities $1/|J_{ss}|$ are not eigenvalues of the
fully coupled chemical Jacobian, and both the numerical value of $\Dafr$ and any threshold
based on it depend on the chosen chemical-time definition and sampling location. Sensitivity
to these choices is examined in \S\ref{sec:sampling} and in the supplementary material, while
the definition above is used consistently throughout the main analysis.

\subsection{Parameter space and sampled conditions}
\label{sec:database}

The controlled chemistry study is evaluated over two sets of edge thermal states selected to
span a broad range of hypersonic wall-layer conditions. The first set uses a cold flight-edge
temperature, $T_e=297.1$~K, while the second samples hotter edge states with $T_e=1000$,
$1350$, $1755$, $2100$ and $2500$~K. Both sets use the same prescribed air composition at the
matching height given in \S\ref{sec:wallmodel}, so differences between them arise from the
imposed thermodynamic state rather than from changes in edge composition.

For the cold-edge conditions the parameter space spans $Ma_e = 3,4,\ldots,25$ and
$p_e\in\{2,5,10,15,20,30,50,70,100\}$~kPa. A common wall temperature cannot be imposed across
this range: at $Ma_e=5$, for example, a $2000$~K wall exceeds the recovery temperature and
reverses the heat flux. The wall temperature is therefore specified relative to an empirical
cold-wall reference correlation,
\begin{equation}
T_{w,\mathrm{ref}}(Ma_e,p_e)=121.15\,Ma_e^{0.7153}\,p_e^{0.0933}~\mathrm{K},
\qquad p_e\text{ in Pa},
\label{eq:twref}
\end{equation}
which gives $1180$~K at $Ma_e=5$, $p_e=20$~kPa and $3320$~K at $Ma_e=25$, $p_e=100$~kPa, with
$T_w/T_{w,\mathrm{ref}}\in\{0.8,0.9,1.0,1.1,1.2\}$. This normalisation maintains physically
comparable wall-cooling levels across conditions with widely different recovery temperatures.

The hot-edge set spans
\[
\begin{gathered}
Ma_e\in\{3,4,5,6.5,8,9,10,11,12\},\qquad T_w\in\{1500,2500\}~\mathrm{K},\\
p_e\in\{2,3,5,7,10,20,50,100\}~\mathrm{kPa},
\end{gathered}
\]
and contains every combination of these axes. The cold-edge correlation is not applied to these conditions
because it was calibrated at $T_e=297$~K and is not assumed to remain valid at elevated edge
temperatures.

Together the sampled conditions comprise $450$ cold-edge and $622$ hot-edge states. The
friction Reynolds number and the maximum temperature are defined as
\begin{equation}
Re_\tau=\frac{\rho_wu_\tau\delta_{99}}{\mu_w},
\qquad
T_{\max}=\max_{0\leq y\leq y_m}T(y),
\label{eq:retau_tmax}
\end{equation}
and span $Re_\tau=181$ to $3.1\times10^4$ and $T_{\max}=478$ to $7993$~K, with wall
temperatures from $432$ to $4256$~K. Whenever $T_{\max}$ is used later as a model-selection
variable, it is evaluated from the frozen solution.

Conditions with $T_{\max}\geq8000$~K are excluded because ionisation may no longer be
negligible, whereas the adopted thermochemical model contains only neutral species; describing
those states would require a thermochemical model that accounts for ionisation
\citep{gupta1990review}. The cutoff limits the cold-edge set to $Ma_e\leq25$ and removes $98$
hot-edge conditions with $T_{\max}=8001$--$11563$~K (\S\ref{sec:validity}). The retained
parameter space therefore defines the range over which the present one-temperature,
non-ionising wall-model formulation is applied, and contains $1072$ primary conditions.
Details of the adaptive refinement strategy, the subset sizes associated with the additional
chemistry levels, and the numerical convergence checks are given in the supplementary
material.

\FloatBarrier

\subsection{Grid independence}
\label{sec:grid}

The production grid uses $y_1^+=0.0125$, a stretching ratio of $1.01$ and
$100\leq N\leq10^4$. Halving $y_1^+$ changes $q_w$ by $0.008$--$0.023\,\%$.
Because an integral heat flux can conceal an underresolved reaction layer, 36 conditions
were also recomputed on three grids. Between the two finest grids, the integrated chemical
source differs by $0.20\,\%$ at the ninety-fifth percentile and $0.25\,\%$ at its maximum;
the corresponding flux, source and wall-composition checks are reported in the supplementary
material.

A second check examines the grid sensitivity of the heat-flux differences used in the
selection criterion of \S\ref{sec:map}. Recomputing the eight conditions nearest the $1\,\%$ boundary changes
either difference by at most $0.0010\,\%$ and changes no label. Detailed grid results are given
in the supplementary material.

\section{Frozen-state criteria for finite-rate-chemistry selection}
\label{sec:map}

The first objective is to determine when finite-rate chemistry must be retained to predict the
wall heat flux within a prescribed tolerance, and whether that decision can be made from a
frozen-chemistry solution alone. The retention criterion is therefore defined directly from
differences in the computed wall heat flux, independently of any Damk\"ohler number or
temperature-based predictor. The frozen-state quantities $\Dafr$ and $T_{\max}$ are then
tested against this independently defined response. This separation allows the timescale
diagnostic to be evaluated as a predictor of chemistry sensitivity rather than being built
into its definition.

\subsection{Heat-flux criterion for retaining finite-rate chemistry}
\label{sec:mapdef}

For conditions with positive reference heat flux, the response to changes in reaction rate is
measured relative to the finite-rate solution at $\Gamma=1$ through two one-sided differences,
\begin{equation}
d_{dn}=\frac{|q_w^{\mathrm{frozen}}-q_w^{\Gamma=1}|}{q_w^{\Gamma=1}},\qquad
d_{up}=\frac{|q_w^{\Gamma=100}-q_w^{\Gamma=1}|}{q_w^{\Gamma=1}}.
\label{eq:dndup}
\end{equation}
Here $d_{dn}$ measures the error
introduced by replacing the reference finite-rate solution with frozen chemistry, whereas
$d_{up}$ measures the additional response obtained when the reaction rates are increased by
two decades from the reference mechanism. The two quantities are retained separately because
opposite changes on the two sides of $\Gamma=1$ can nearly cancel in a difference formed only
between the limiting states: in some conditions, oppositely directed one-sided responses of
$2.405$ and $2.419\,\%$ leave a net difference of only $0.01\,\%$.

A prescribed tolerance $\theta$ determines whether frozen chemistry is sufficiently accurate
for the target wall-heat-flux prediction: finite-rate chemistry is retained when
$d_{dn}\geq\theta$, and frozen chemistry is considered adequate when $d_{dn}<\theta$. Unless
stated otherwise the results below use $\theta=1\,\%$. This value is an allowable relative
difference in wall heat flux between the frozen and reference finite-rate solutions; it is not
an estimate of the total uncertainty of the wall model.

The upper-rate response provides a secondary distinction. Combining $d_{dn}$ and $d_{up}$
gives the four classes
\begingroup
\setlength{\abovedisplayskip}{10pt}
\setlength{\belowdisplayskip}{10pt}
\begin{equation}
\setlength{\arraycolsep}{4pt}
\begin{array}{c|cc}
 & d_{up}<\theta & d_{up}\geq\theta \\
\hline
d_{dn}<\theta
  & \begin{array}{c}\text{frozen adequate;}\\\text{upper-rate insensitive}\end{array}
  & \begin{array}{c}\text{frozen adequate;}\\\text{upper-rate sensitive}\end{array} \\[2mm]
d_{dn}\geq\theta
  & \begin{array}{c}\text{finite-rate required;}\\\text{upper-rate insensitive}\end{array}
  & \begin{array}{c}\text{finite-rate required;}\\\text{upper-rate sensitive}\end{array}
\end{array}.
\label{eq:zones}
\end{equation}
\endgroup

The primary model-selection decision therefore depends only on $d_{dn}$; $d_{up}$ determines
whether the wall heat flux remains appreciably sensitive as the reaction rates are increased
beyond their reference values. The construction uses three solutions per condition and
requires neither a fast-chemistry endpoint nor a Damk\"ohler-number input.

\begin{figure}[!htbp]
  \centering
  \begin{subfigure}{0.44\textwidth}\caption{}%
    \includegraphics[width=\textwidth]{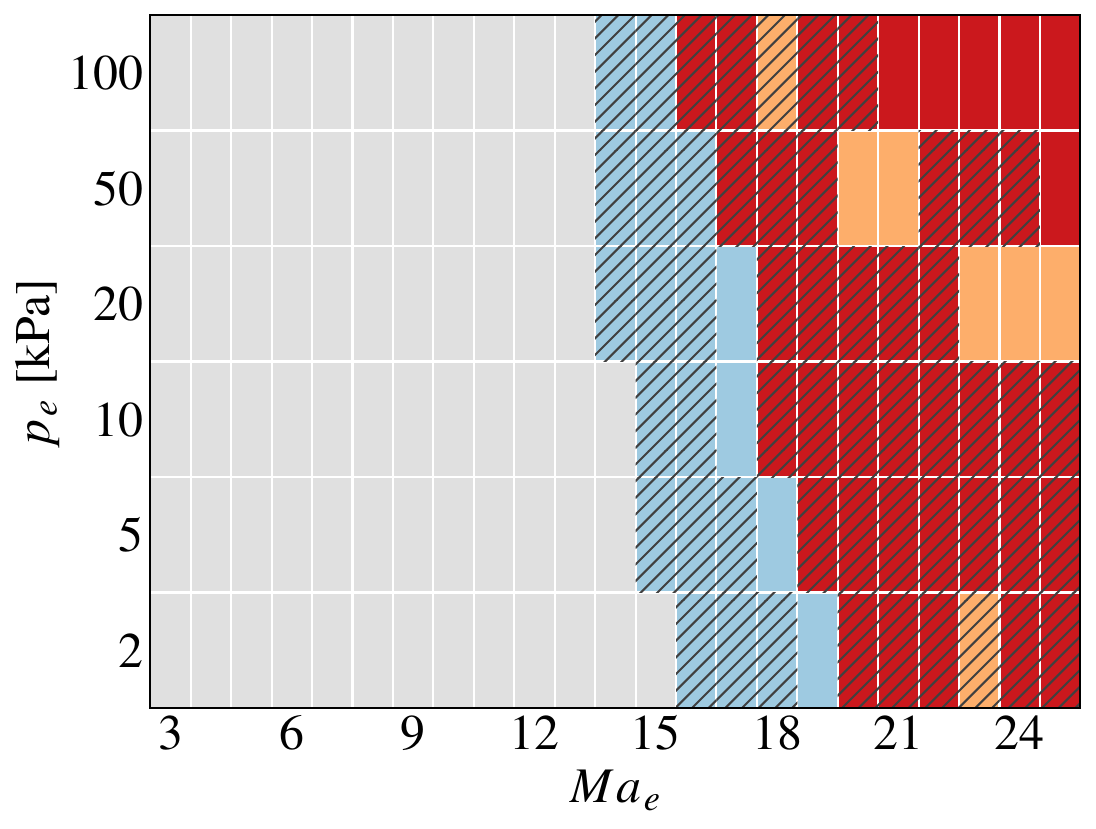}%
    \label{fig:zoning_a}\end{subfigure}\hfill
  \begin{subfigure}{0.44\textwidth}\caption{}%
    \includegraphics[width=\textwidth]{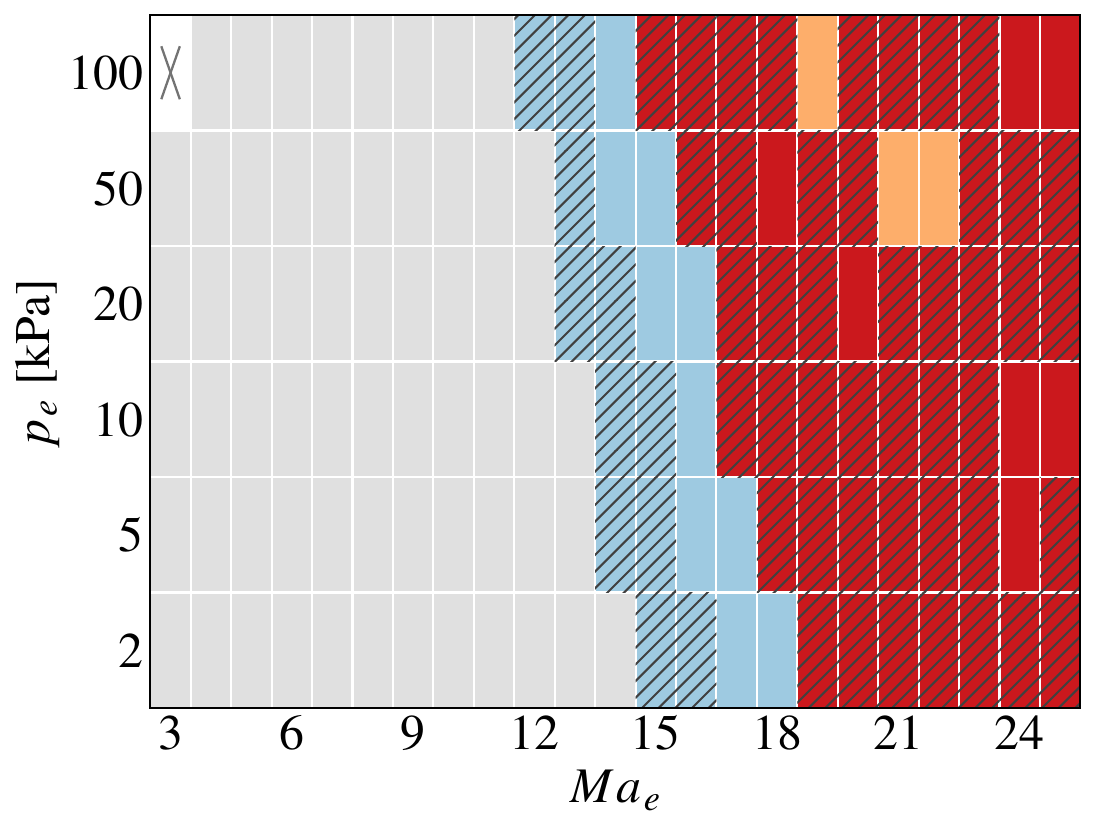}%
    \label{fig:zoning_b}\end{subfigure}

  \begin{subfigure}{0.44\textwidth}\caption{}%
    \includegraphics[width=\textwidth]{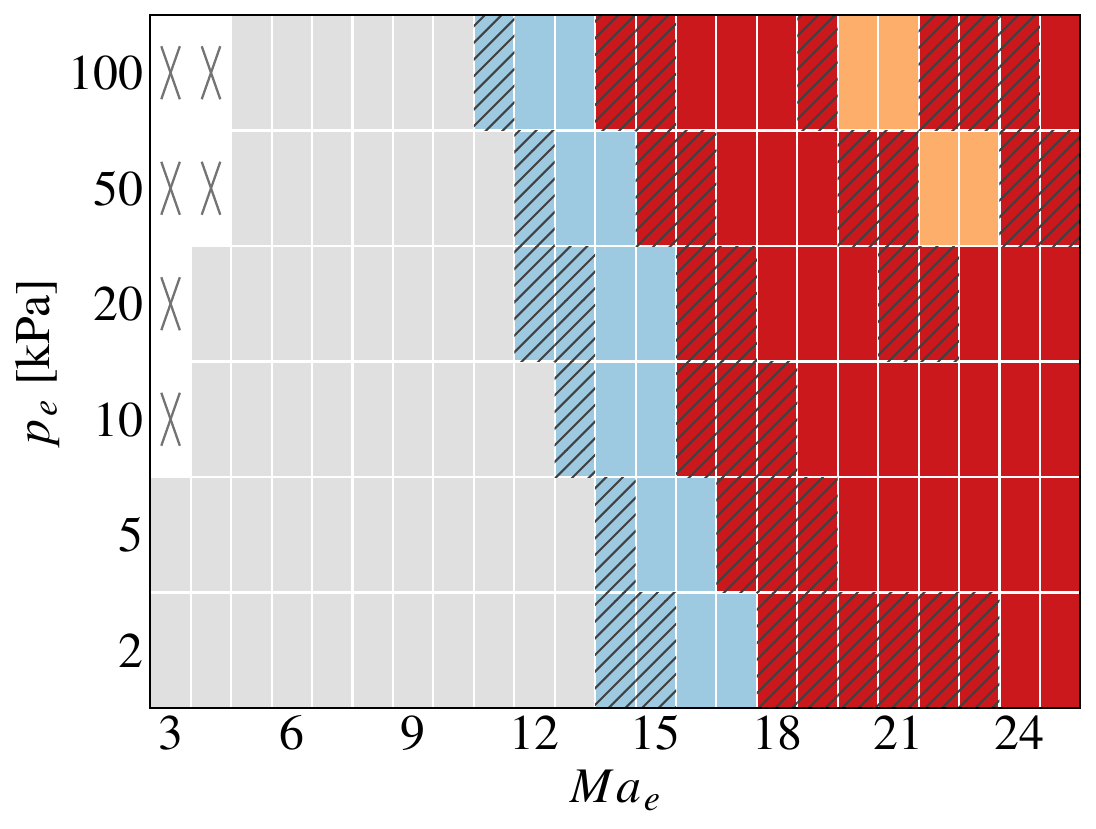}%
    \label{fig:zoning_c}\end{subfigure}\hfill
  \begin{subfigure}{0.44\textwidth}\caption{}%
    \includegraphics[width=\textwidth]{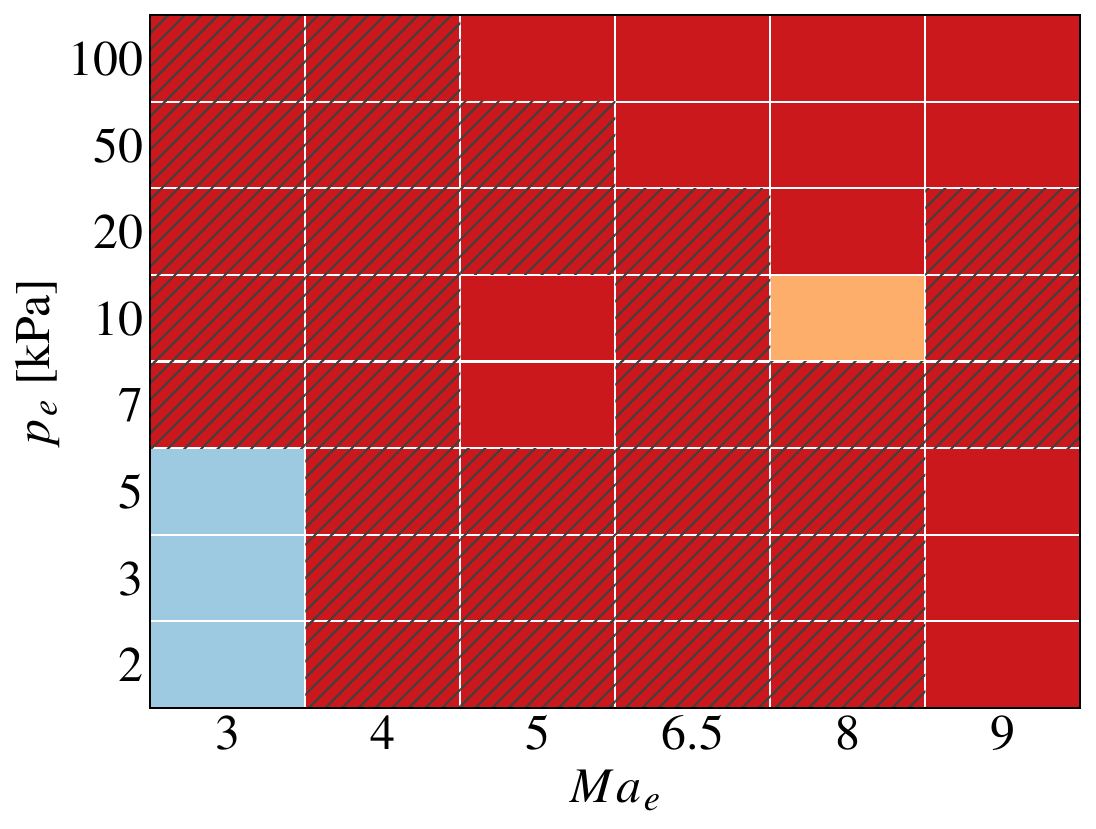}%
    \label{fig:zoning_d}\end{subfigure}

  \includegraphics[width=0.88\textwidth]{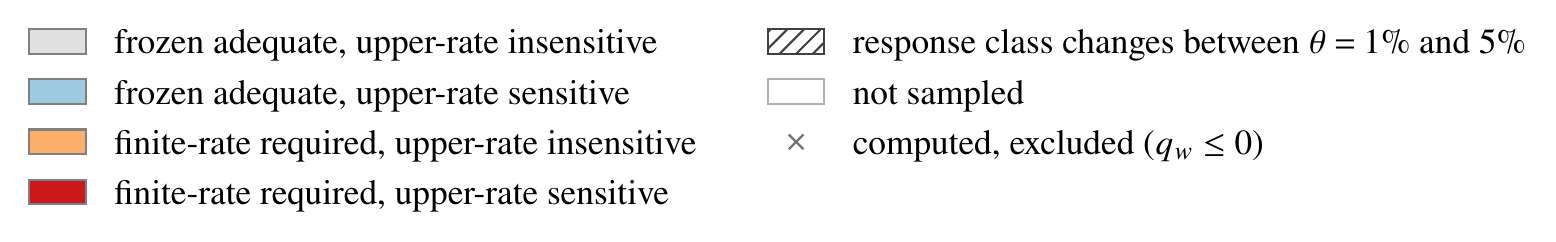}
  \caption{Rate-sensitivity map over $(Ma_e,p_e)$. (\textit{a})--(\textit{c}) cold-edge set
  at wall-temperature ratios $T_w/T_{w,\mathrm{ref}}=0.8$, $1.0$ and $1.2$;
  (\textit{d}) hot-edge set at $T_e=T_w=2500$~K. Background colours show the four response
  classes of \eqref{eq:zones} at the operating tolerance $\theta=1\,\%$. Hatching marks the 175
  cells whose class differs between
  $\theta=1\,\%$ and $\theta=5\,\%$. Of the 455 coloured cells, 407 belong to the three
  cold-edge wall-temperature-ratio panels and 48 to the plotted hot-edge slice. Crosses
  mark seven computed cells excluded because $q_w\le0$; blank cells were not sampled. In the
  cold-edge set, the nearly vertical boundary moves from $Ma_e\simeq15$ at $\theta=1\,\%$ to
  $Ma_e\simeq20$ at $\theta=5\,\%$. The hot-edge set retains finite-rate chemistry over
  most of its range at $\theta=1\,\%$ despite its lower Mach numbers, so $Ma_e$ alone cannot
  define the boundary across both families.}
  \label{fig:zoning}
\end{figure}

Figure~\ref{fig:zoning} shows how this heat-flux criterion varies across the sampled parameter
space. For the cold-edge conditions at $T_w/T_{w,\mathrm{ref}}=1$, finite-rate chemistry is
not required below approximately $Ma_e=15$ at the $1\,\%$ tolerance. The fraction of
conditions requiring finite-rate chemistry then increases rapidly between approximately
$Ma_e=15$ and $18$ and reaches all sampled pressures by $Ma_e=20$, so the transition occupies
a finite range of Mach number rather than a single critical value. More importantly, many
hot-edge conditions require finite-rate chemistry at substantially lower Mach numbers. The
selection boundary is consequently governed by the thermochemical state rather than by Mach
number alone.

Increasing the allowable heat-flux difference moves the retention boundary towards more
strongly reacting conditions, so the thresholds reported below are associated with the stated
value of $\theta$ and must be recalibrated when the tolerance is changed. Results for
$\theta=0.5$, $2$ and $5\,\%$, together with the sensitivity to the upper-rate probe, are
given in the supplementary material.

\subsection{Frozen-state criteria for the retention decision}
\label{sec:criteria}

Two quantities available from the frozen solution are considered as practical indicators of
the finite-rate-retention boundary: the timescale diagnostic $\Dafr=t_{shear}/t_{ch}$ defined
in \S\ref{sec:da}, and the maximum temperature $T_{\max}$ of the frozen wall-layer solution.
The former expresses the competition between local chemical and shear timescales, whereas the
latter provides a directly available thermochemical state variable without requiring
evaluation of the chemical Jacobian.

For the $1\,\%$ wall-heat-flux tolerance the fitted frozen-state criteria are
\begin{equation}
\Dafr\geq1.93\times10^{3},
\qquad
T_{\max}\geq3760~\mathrm{K}.
\label{eq:primary_thresholds}
\end{equation}
Across the $1057$ conditions with positive reference heat flux, of which $573$ satisfy
$d_{dn}\geq1\,\%$, these thresholds give balanced accuracies of $96.4$ and $94.7\,\%$.
Balanced accuracy is the mean of the true-positive and true-negative rates and is used because
the two retention classes are not exactly equal in size.
\begin{table}[htbp]
  \centering
  \footnotesize
  \setlength{\tabcolsep}{3pt}%
  \begin{tabular}{lcccc}
  \hline\hline
   & Full-fit & Random five-fold & \multicolumn{2}{c}{Cross-family balanced accuracy (FNR)} \\
  Predictor & threshold; BA & BA & cold-edge $\to$ hot-edge & hot-edge $\to$ cold-edge \\
  \hline
  $\Dafr$ & $1.93\times10^3$; $96.4\,\%$ & $95.8\,\%$ & $94.6\,\%$ ($8.8\,\%$) & $97.6\,\%$ ($0.0\,\%$) \\
  $T_{\max}$ & $3760$~K; $94.7\,\%$ & $94.5\,\%$ & $91.4\,\%$ ($14.7\,\%$) & $94.7\,\%$ ($0.0\,\%$) \\
  \hline\hline
  \end{tabular}
  \caption{Performance of the binary decision to retain finite-rate chemistry. All $1057$
  conditions with positive reference heat flux are used; $573$ satisfy
  $d_{dn}\geq1\,\%$ and therefore require finite-rate chemistry. Thresholds are selected by ordinary
  accuracy on the calibration set. BA denotes balanced accuracy and FNR the false-negative
  rate. The corresponding full-fit accuracies are $96.5$ and $94.6\,\%$, and the areas under
  the receiver-operating-characteristic curve are $0.996$ and $0.989$.}
\label{tab:criteria}
\end{table}

The criteria remain effective when entire Mach-number or pressure slices are withheld during
calibration. The corresponding pooled balanced accuracies are $95.5$ and $95.9\,\%$ for
$\Dafr$ and $93.1$ and $94.2\,\%$ for $T_{\max}$. Imposing a training-side
false-negative-rate limit of $5\,\%$ gives held-out false-negative rates of
$4.4$--$5.1\,\%$. Cross-family transfer is more demanding, particularly when criteria
calibrated on the cold-edge conditions are applied to the hot-edge set, but the classification
remains above $90\,\%$ balanced accuracy. These tests indicate that the criteria are not
simply reproducing a local Mach- or pressure-dependent boundary.

A separate test examines sensitivity to the matching-height scale without refitting either
threshold. Thirty-six additional conditions were evaluated at reference streamwise scales of
$x_{\mathrm{ref}}=0.5$ and $2$~m. The fixed criterion $\Dafr=1.93\times10^{3}$ gives a
balanced accuracy of $90.9\,\%$ with no false negatives, whereas the fixed
$T_{\max}=3760$~K criterion gives $98.0\,\%$ with one false negative. The numerical thresholds
should nevertheless be interpreted as model- and tolerance-specific values rather than
universal critical constants.
\begin{figure}[htbp]
  \centering
  \begin{subfigure}{0.48\textwidth}\caption{}%
    \includegraphics[width=\textwidth]{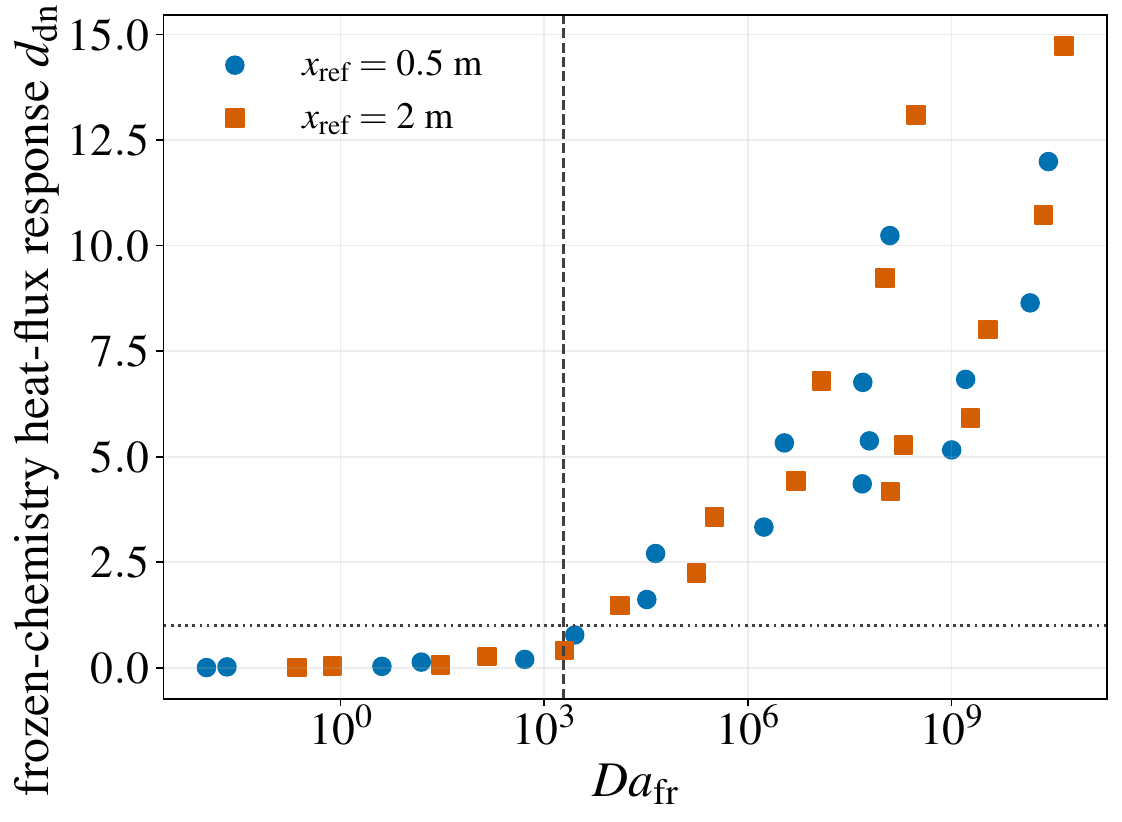}%
    \label{fig:blind-xref-da}\end{subfigure}\hfill
  \begin{subfigure}{0.48\textwidth}\caption{}%
    \includegraphics[width=\textwidth]{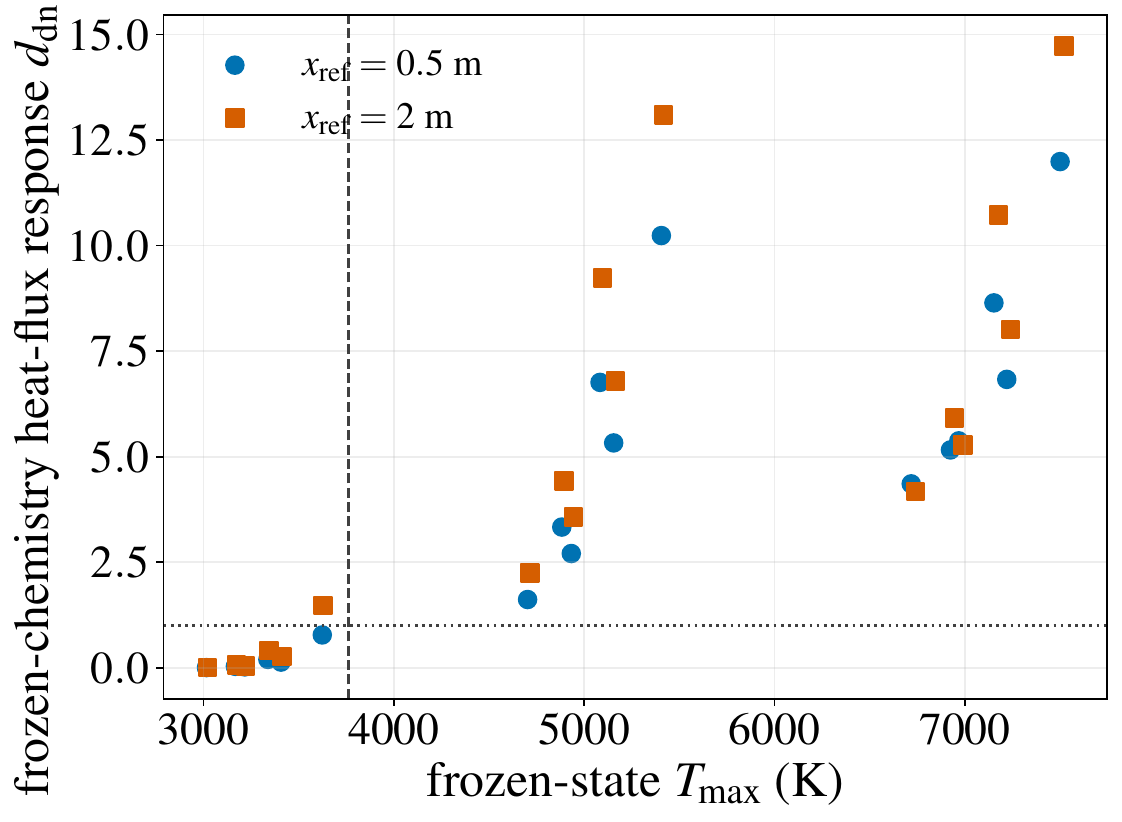}%
    \label{fig:blind-xref-tmax}\end{subfigure}
  \caption{Application of the locked primary criteria to the off-grid matching-height test set.
  The horizontal dotted line is the $1\,\%$ retention tolerance and the vertical dashed lines
  are the previously calibrated thresholds: (\textit{a}) $\Dafr=1.93\times10^3$ and
  (\textit{b}) $T_{\max}=3760$~K. Circles and squares denote the paired
  $x_{\mathrm{ref}}=0.5$ and $2$~m calculations. These tests vary the matching-height scale
  within the same wall model and do not constitute external reacting-flow validation.}
  \label{fig:blind-xref}
\end{figure}

The two criteria provide complementary information. $T_{\max}$ is simple to evaluate and
closely follows the onset of strong high-temperature chemistry, whereas $\Dafr$ retains an
explicit comparison between chemical and wall-layer timescales and therefore gives a more
direct physical interpretation of why the finite-rate contribution becomes important.

The remaining disagreements occur within the boundary band and reveal a genuine difference
between the predictors. Along the $77$ hot-edge sequences in which only pressure is varied,
\begin{equation}
\Dafr \propto p_e^{\,k},
\label{eq:dap}
\end{equation}
with a median fitted exponent of $1.53$ (10th--90th percentiles:
$1.13$--$2.54$), while the median within-sequence variation of $T_{\max}$ is $0.7\,\%$
(maximum $1.3\,\%$). 
Pressure therefore produces a much stronger variation in $Da_{\mathrm{fr}}$
than in $T_{\max}$ along these sequences. Using one threshold shared across the two families reduces the
accuracy by less than one percentage point for either criterion. The role of this residual pressure dependence in the momentum-side relation
is examined in \S\ref{sec:scaling}.

\subsection{Continued rate sensitivity and chemical interpretation}
\label{sec:whytmax}

The primary criterion answers whether frozen chemistry is sufficiently accurate at the
reference reaction rates. A separate question is whether the wall heat flux continues to
change appreciably when the rates are increased from $\Gamma=1$ to $\Gamma=100$, which
corresponds to $d_{up}\geq\theta$ among the conditions for which finite-rate chemistry is
retained. For this continued sensitivity at $\theta=1\,\%$, frozen-state thresholds of
approximately $\Dafr\gtrsim4.9\times10^3$ and $T_{\max}\gtrsim3700$~K give balanced accuracies
of $91.4$ and $89.5\,\%$. These values describe the response to accelerated kinetics and
should not be confused with the primary retention thresholds in
\eqref{eq:primary_thresholds}.

The chemical-time diagnostic also indicates which processes are associated with this
transition. In the diagonal-Jacobian definition of $t_{ch}$, N$_2$ supplies the longest
active-species timescale in $95.8\,\%$ of the diagnosed conditions and in every condition
exhibiting a measurable chemistry response. The corresponding apparent activation-energy
range, $915$--$1052$~kJ\,mol$^{-1}$, contains the $941$~kJ\,mol$^{-1}$ value of the N$_2$
dissociation reaction in the adopted mechanism. These observations associate the
frozen-to-finite-rate transition with the slow nitrogen chemistry represented by the
mechanism, while not implying that a single reaction uniquely controls the full coupled
wall-layer response. A Jacobian time also depends on the local composition and third-body
concentrations, so the agreement is suggestive rather than conclusive. Species-resolved and
pressure-dependent results are given in the supplementary material.

\subsection{Dependence on the sampling location}
\label{sec:sampling}

Because $\Dafr$ is a local wall-layer diagnostic, its numerical value depends on where the
chemical and shear timescales are sampled. Evaluating the diagnostic at $y^+=50$, $100$ and
$200$ gives similar performance for identifying the continued rate sensitivity, with balanced
accuracies between $90.5$ and $91.2\,\%$, but the fitted numerical thresholds differ by
approximately a factor of five. The classification is therefore substantially more robust than
the numerical value of the threshold itself.

A second comparison replaces the fixed $y^+=100$ value by the maximum $\Dafr$ over the fitted
logarithmic-layer interval. The two definitions give different classifications in only $9$ of
$1037$ conditions, even though their condition-by-condition ratio spans nearly an order of
magnitude. This supports the use of $y^+=100$ as a practical sampling location within the
present wall model, but it does not make the resulting threshold universal. The fitted value
in \eqref{eq:primary_thresholds} is specific to the chemical-time definition, sampling
convention, wall model and heat-flux tolerance used here. Detailed comparisons of alternative
sampling locations and chemical-time definitions are reported in the supplementary material.

\section{Wall-heat-flux errors and non-monotone reaction-rate response}
\label{sec:cost}

The criteria in \S\ref{sec:map} identify when frozen chemistry no longer reproduces the
reference finite-rate wall heat flux within the prescribed tolerance. The next question is
what is lost when the finite-rate solution is replaced by either limiting chemistry treatment.
This section compares the wall-heat-flux errors of the frozen and infinitely fast
reaction-rate limits and then examines the response between them. The results show that the
two limits are not equivalent approximations to finite-rate chemistry and, more importantly,
that the finite-rate wall heat flux does not generally vary monotonically between them.

\subsection{Errors of the limiting chemistry treatments}
\label{sec:limiterr}

The frozen and high-rate limits are compared with the reference finite-rate solution at
$\Gamma=1$ using the same prescribed matching-edge state. The frozen-chemistry error is
therefore identical to $d_{dn}$ in \eqref{eq:dndup}, while the high-rate error is evaluated
from the $\Gamma\to\infty$ approximation described in \S\ref{sec:gamma}. The latter is the
infinitely fast reaction-rate limit of the same fixed-edge-state boundary-value problem; it
should not be interpreted as a separate equilibrium closure in which the matching-edge
composition is recomputed.

\begin{figure}[H]
  \centering
  \begin{subfigure}{0.48\textwidth}\caption{}%
    \includegraphics[width=\textwidth]{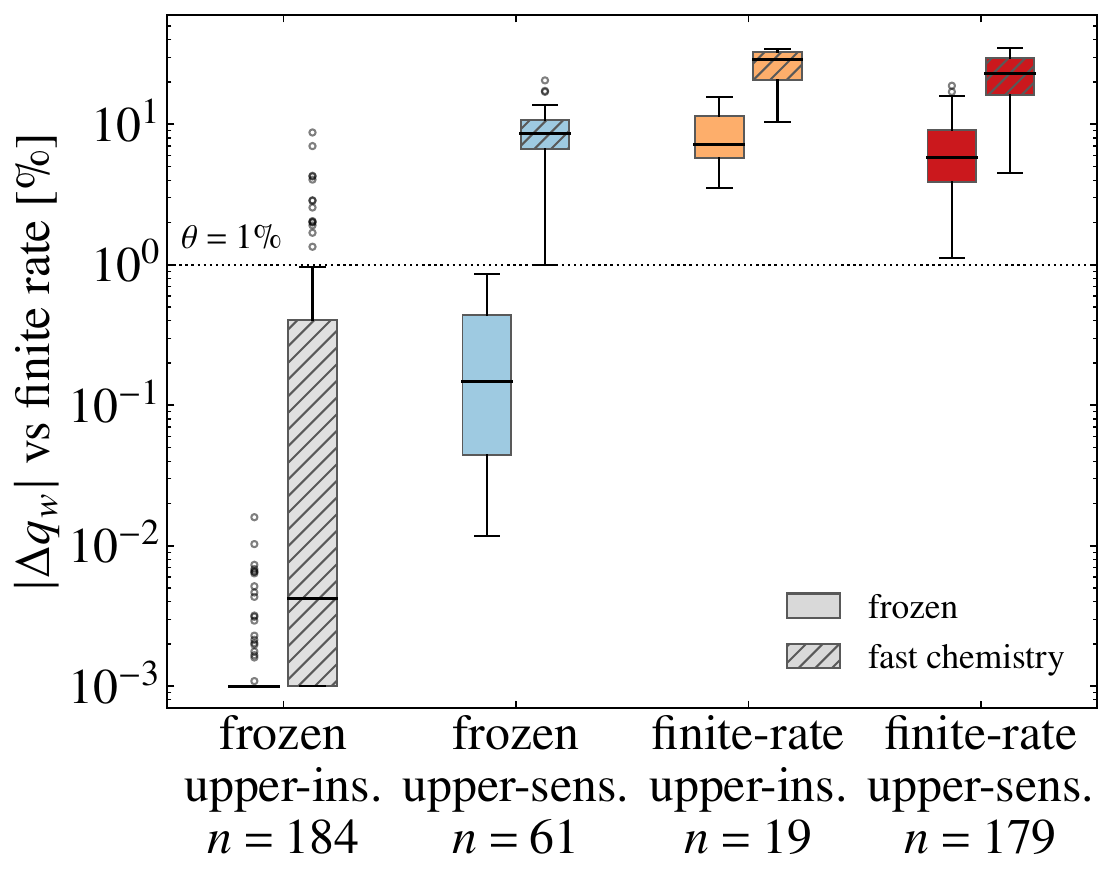}%
    \label{fig:costzone_a}\end{subfigure}\hfill
  \begin{subfigure}{0.48\textwidth}\caption{}%
    \includegraphics[width=\textwidth]{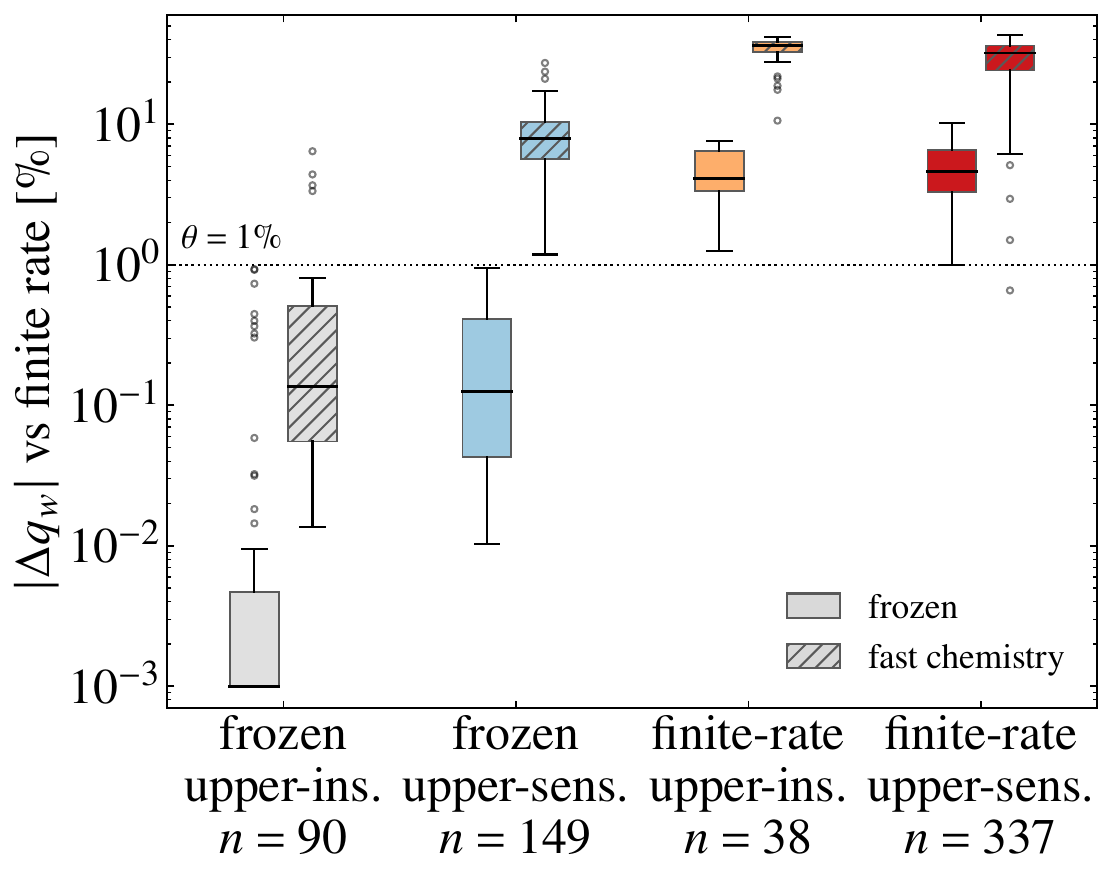}%
    \label{fig:costzone_b}\end{subfigure}
  \caption{Distribution of the $q_w$ error incurred by each limiting chemistry treatment, resolved by
  response class: (\textit{a}) cold-edge set and (\textit{b}) hot-edge set. Solid
  boxes denote frozen chemistry and hatched boxes the fast-chemistry approximation. Frozen-chemistry errors
  in the first two classes are small by construction; the remaining contrasts are not imposed by
  the class definitions. In both classes for which finite-rate chemistry is required, a fast-chemistry tier
  is available for every condition.}
  \label{fig:costzone}
\end{figure}

Figure~\ref{fig:costzone} compares the two errors for the four response classes defined in
\eqref{eq:zones}. Frozen-chemistry errors are necessarily below the prescribed tolerance in
the two classes for which frozen chemistry is adequate, but the high-rate errors are not
constrained by the classification. Where finite-rate chemistry is required, the two limiting
treatments behave very differently. In particular, conditions that become only weakly
sensitive when the rate is increased from $\Gamma=1$ to $\Gamma=100$ can nevertheless remain
far from the infinitely fast reaction-rate limit. Weak local sensitivity to a further increase
in reaction rate therefore does not imply proximity to chemical equilibrium.

The difference between the two limiting approximations becomes especially large at high Mach
number. Across the cold-edge conditions, the frozen-chemistry error remains below
$0.12\,\%$ for $Ma_e\leq12$, whereas the high-rate error reaches $2.9\,\%$. For the $144$
sampled conditions with $Ma_e\geq20$, the median wall-heat-flux errors are $7.1\,\%$ for
frozen chemistry and $27.6\,\%$ for the infinitely fast reaction-rate approximation, with
non-overlapping interquartile ranges. Across the complete database, the ninety-fifth
percentile and maximum errors are $9.2$ and $18.8\,\%$ for frozen chemistry, compared with
$39.2$ and $43.2\,\%$ for the high-rate limit.

This ordering is also evident after conditioning on the finite-rate-retention criterion. In
the conditions that are both finite-rate-required and upper-rate-sensitive, the median
frozen-chemistry error is $5.8\,\%$ in the cold-edge set and $4.5\,\%$ in the hot-edge set. By
contrast, in the finite-rate-required but upper-rate-insensitive class, the median error of
the infinitely fast reaction-rate limit is $29.0$ and $36.1\,\%$ in the two sets. The
high-rate limit can therefore remain a poor representation of the reference finite-rate heat
flux even after the response between $\Gamma=1$ and $\Gamma=100$ has become small.

These results show that frozen and infinitely fast chemistry are not interchangeable limiting
approximations. Under the present fixed-edge conditions, frozen chemistry is often the closer
of the two limits at high Mach number, consistent with a strongly cooled near-wall region
remaining thermochemically far from the infinitely fast state.

\subsection{Non-monotone response to reaction rate}
\label{sec:nonmono}

The large difference between the two limiting approximations raises a more fundamental
question: does the finite-rate wall heat flux at least remain between the frozen and
infinitely fast reaction-rate values? The calculations show that it generally does not.

Among the $778$ positive-heat-flux conditions for which the high-rate endpoint is available,
the reference finite-rate value $q_w(\Gamma=1)$ lies more than $1\,\%$ below both limiting
values in $572$ cases, or $73.5\,\%$ of the sample. This behaviour occurs in both
thermal-state sets and is therefore not restricted to one part of the parameter space. The
finite-rate solution at the unmodified reaction rates can consequently fall outside the range
defined by the frozen and infinitely fast reaction-rate predictions.

\begin{figure}[H]
  \centering
  \includegraphics[width=0.6\textwidth]{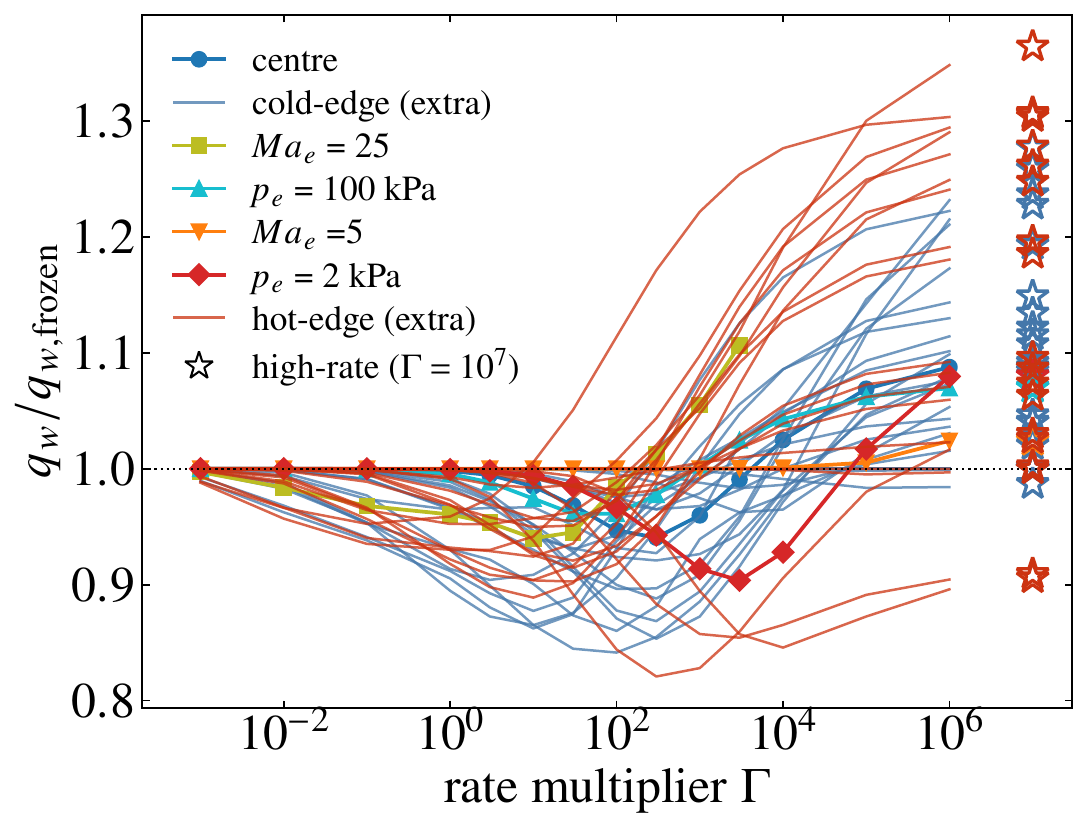}
  \caption{$q_w$ normalised by its frozen value along the rate axis. Symbols mark the five
  reference conditions, thin lines the forty-five stratified conditions and stars the
  fast-chemistry tier. All reference conditions have $T_e=297.1$~K and $T_w=2000$~K: the centre
  is $(Ma_e,p_e)=(15,20~\mathrm{kPa})$, the low/high-Mach cases use
  $(5/25,20~\mathrm{kPa})$, and the low/high-pressure cases use
  $(15,2/100~\mathrm{kPa})$. The stars are placed at the computed value $\Gamma=10^{7}$ rather than at
  the right-hand edge. Forty-two of the fifty curves fall more than $0.1\,\%$ below the frozen
  value; all forty-one of these curves with an available fast-chemistry endpoint also fall below
  that endpoint. The $0.1\,\%$ level is used only to identify a resolved undershoot on a rate
  ladder and is distinct from the $1\,\%$ model-selection tolerance. The remaining conditions are weakly affected by chemistry over the
  finite-rate range. The minimum occurs between $\Gamma=0.1$ and $10^{5}$, which is not
  sampled by a two-point endpoint calculation.}
  \label{fig:nonmono}
\end{figure}

To determine whether this reflects an isolated comparison at $\Gamma=1$ or a broader
dependence on reaction rate, the response was resolved using fourteen values of $\Gamma$ for
$50$ representative conditions spanning the sampled thermodynamic states and
chemistry-response classes. Figure~\ref{fig:nonmono} shows the resulting wall-heat-flux
trajectories. In $42$ of the $50$ conditions, $q_w$ first decreases below its frozen value as
the reaction rates increase and subsequently rises at larger $\Gamma$. For the $41$ of these
conditions with a converged high-rate endpoint, the internal minimum also lies below the
infinitely fast reaction-rate value.

The depth and position of this minimum vary substantially among conditions. For the $42$
non-monotone responses, the minimum heat flux ranges from $0.821$ to $0.998$ of the frozen
value, with a median deficit of $7.4\,\%$, and the corresponding minima occur over
$10^{-1}\lesssim\Gamma\lesssim10^{5}$, spanning approximately six decades of reaction-rate
multiplier. A single intermediate value of $\Gamma$ therefore cannot locate the minimum
consistently across the parameter space.

The physical consequence matters for model interpretation. Increasing $\Gamma$ does not move $q_w$ monotonically from the frozen value towards the infinitely fast reaction-rate limit. The coupled changes in composition,
thermal state, species-enthalpy transport and wall-layer structure can instead reduce $q_w$
before the high-rate asymptote is approached. The finite-rate response is therefore not, in
general, an interpolation between its two limiting chemistry solutions, and knowledge of the
endpoints alone is insufficient to determine either the magnitude or even the direction of the
intermediate wall-heat-flux response.

\subsection{Approach to the infinitely fast reaction-rate limit}
\label{sec:eqlimit}

Because the upper endpoint is obtained at a large but finite value of $\Gamma$, its proximity
to the $\Gamma\to\infty$ limit must be established before the endpoint comparisons above can
be interpreted. The high-rate state differs conceptually from an algebraically imposed
equilibrium solution: the matching-edge composition remains fixed as $\Gamma$ increases, so a
thin reaction--diffusion layer can persist near the edge or wall to reconcile the imposed
boundary composition with rapid chemical relaxation in the interior.

The asymptotic approach was examined for $20$ conditions using additional solutions over
$10^{6}\leq\Gamma\leq10^{9}$. Nineteen conditions completed the full high-rate sequence, and
in $18$ of these the magnitude of the per-decade increment in $\ln q_w$ decreases
monotonically as $\Gamma$ increases, indicating approach towards an asymptotic
wall-heat-flux value.

For a non-catalytic boundary, a local balance between diffusion and a reaction rate
proportional to $\Gamma$ gives a reaction--diffusion thickness
\begin{equation}
\delta_{\mathrm{rd}}\sim\Gamma^{-1/2}.
\label{eq:drd}
\end{equation}
If the leading high-rate correction to $q_w$ scales with this layer thickness, successive
corrections separated by one decade in $\Gamma$ should contract by approximately
$\sqrt{10}=3.162$. The measured contraction factors have a median of $3.166$, and $15$ of the
$19$ complete sequences lie between $2.0$ and $4.0$, so the observed high-rate behaviour is
consistent with the expected thinning of a reaction--diffusion layer.

The remaining distance to the infinite-rate limit can be estimated from the contracting
sequence. If $\delta_{\mathrm{last}}$ is the magnitude of the last per-decade increment and
$r>1$ its measured contraction factor, the remaining geometric tail is
\begin{equation}
\epsilon_\infty=\frac{\delta_{\mathrm{last}}}{r-1}.
\label{eq:drdplus}
\end{equation}
For the complete sequences the estimated residual wall-heat-flux difference is below
$0.61\,\%$, with a median of $0.047\,\%$. Grid refinement of the highest-rate solutions changes
$q_w$ by at most $0.024\,\%$, confirming that the observed contraction is not produced by
progressive loss of wall-normal resolution.

The principal conclusion concerning the non-monotone response is insensitive to this residual
endpoint uncertainty. Even if every computed high-rate heat flux is shifted by $0.61\,\%$ in
the direction most favourable to placing the finite-rate solution between the two limits, the
number of conditions for which $q_w(\Gamma=1)$ lies more than $1\,\%$ below both limiting
values decreases only from $572$ to $571$. The failure of the two limiting predictions to
bound the reference finite-rate wall heat flux is therefore not an artefact of incomplete
convergence of the high-rate endpoint.

The results of this section establish two distinct consequences of finite-rate chemistry.
First, replacing the reference kinetics by either limiting treatment can produce substantial
wall-heat-flux errors, and the infinitely fast reaction-rate limit is not generally the more
accurate approximation. Second, the wall heat flux itself can vary non-monotonically with
reaction rate, so the two limiting solutions do not determine the intermediate finite-rate
response. The following section examines why these substantial changes in wall thermodynamics
and heat transfer produce a much smaller response in the Reynolds analogy factor.

\section{Chemistry-induced response of the Reynolds analogy factor}
\label{sec:hierarchy}

Section~\ref{sec:cost} shows that changing the chemical timescale can alter the wall heat flux
substantially and can produce a strongly non-monotone response between the frozen and
infinitely fast reaction-rate limits. The corresponding Reynolds analogy factor (RAF) is much
less sensitive. Under the baseline transport closure, changes in wall thermodynamic state and
individual transfer quantities can reach tens of per cent, whereas the RAF changes by only a
few per cent. This section examines why the chemistry-induced changes in $St$ and $C_f$ largely
cancel in the RAF and what produces the remaining RAF response.

\subsection{From wall-state changes to the Reynolds analogy factor}
\label{sec:raf_response}

The Reynolds analogy factor $\RAF=2St/C_f$ responds to chemistry not only through the changes
in wall heat flux and wall shear stress, but also through the thermodynamic quantities used to
normalise those fluxes.

For the frozen-to-$\Gamma{=}100$ comparison, the wall density and the enthalpy driving
potential change by as much as approximately $25\,\%$, while the wall heat flux and wall shear
stress change by as much as $19.5$ and $15.6\,\%$. The two dimensional wall fluxes do not
generally respond in the same direction: among conditions with a chemically significant change
in Stanton number, $|\delta\ln St|>2\,\%$, the increments of $q_w$ and $\tau_w$ have the same
sign in only $12.6\,\%$ of cases.
\begin{figure}[H]
  \centering
  \begin{subfigure}{0.49\textwidth}
    \caption{}
    \includegraphics[width=\linewidth]{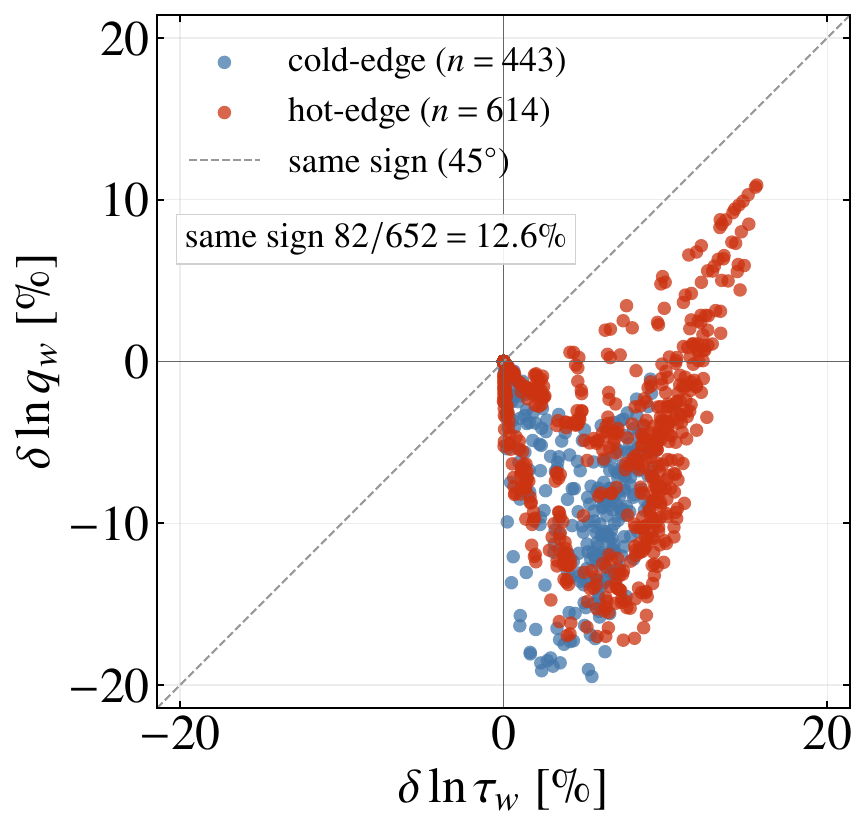}
    \label{fig:flux}
  \end{subfigure}
  \begin{subfigure}{0.49\textwidth}
    \caption{}
    \includegraphics[width=\linewidth]{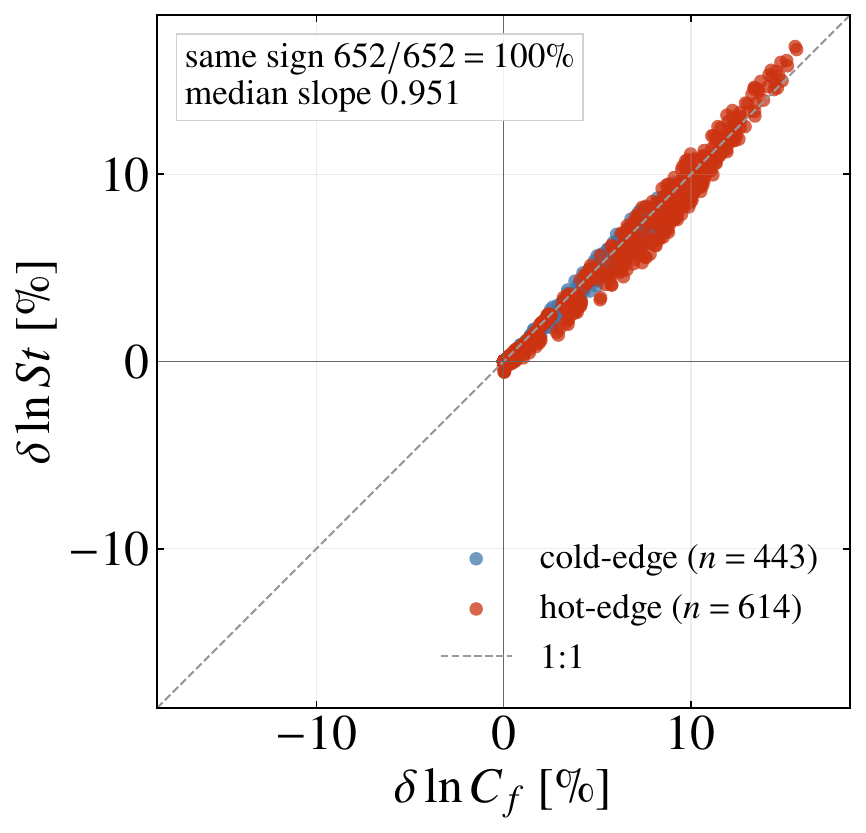}
    \label{fig:coef}
  \end{subfigure}
  \caption{Frozen-to-$\Gamma{=}100$ increments for all $1057$ paired conditions
  ($443$ cold-edge and $614$ hot-edge). The annotations apply the common chemically
  significant filter $|\delta\ln St|>2\,\%$ ($n=652$): (\textit{a}) $q_w$ and $\tau_w$
  have the same sign in only $12.6\,\%$; (\textit{b}) $St$ and $C_f$ have the same sign in
  all cases, with median increment ratio $0.951$. The points near zero that do not pass the
  filter remain visible but do not enter those statistics.}
  \label{fig:fluxcoef}
\end{figure}

After normalisation the behaviour changes markedly. The Stanton number and the half
skin-friction coefficient move in the same direction in all chemically significant
frozen-to-$\Gamma{=}100$ cases, and their logarithmic increments are closely matched, with a
median ratio $\delta\ln St/\delta\ln(C_f/2)=0.951$. Much of the chemistry-induced change
common to the two transfer coefficients therefore cancels when their ratio is formed.

This progression is summarised in table~\ref{tab:hierarchy}. The largest sampled changes
decrease from approximately $25\,\%$ in the wall state and approximately $20\,\%$ in the
individual wall fluxes to $16$--$17\,\%$ in the transfer coefficients, and finally to
$2.12\,\%$ in the RAF over the frozen-to-$\Gamma{=}100$ interval. Over the wider
frozen-to-fast-chemistry interval the largest sampled RAF change is $4.31\,\%$.
\begin{table}
  \centering
  \footnotesize
  \setlength{\tabcolsep}{3pt}%
  \begin{tabular}{llcl}
  \hline\hline
  Level & Quantity & Max.\ change & Key statistic \\
  \hline
  Wall state & $\rho_w$, $\Delta h$ & $25.0\,\%$ & Pearson corr.\ $0.834$ ($n=965$) \\
  Wall fluxes & $q_w$, $\tau_w$ & $19.5$, $15.6\,\%$ & same-sign $12.6$/$97.2\,\%$ \\
  Coefficients & $St$, $C_f/2$ & $16.8$, $15.6\,\%$ & same-sign $100\,\%$; ratio $0.951$ \\
  Their ratio & $\RAF=2St/C_f$ & $2.12$ / $4.31\,\%$ & see text for the two pairings \\
  \hline\hline
  \end{tabular}
  \caption{Response to chemistry at four levels. The wall-state,
  flux and coefficient maxima use the $965$ transitional pairs for which $\RAF$ is defined.
  Same-sign statistics use the chemically significant subset $|\delta\ln St|>2\,\%$: $652$
  of $1057$ transitional pairs and $650$ of $778$ full-span pairs. The two same-sign percentages in the wall-flux row correspond to the
transitional and full-span chemically significant subsets, respectively. The coefficient
  row reports the transitional same-sign rate and median $\delta\ln St/\delta\ln C_f$; the
  corresponding full-span values are $99.8\,\%$ and $1.170$. The two $\RAF$ values use the
  common intersection of $712$ conditions ($281$ cold-edge and $431$ hot-edge) on which both
  pairings are valid, as does table~\ref{tab:decomp}. The reported correlation is the Pearson
  correlation between $\delta\ln\rho_w$ and $\delta\ln\Delta h$.}
  \label{tab:hierarchy}
\end{table}

The small RAF change therefore does not imply weak chemistry effects on the
underlying wall state or wall fluxes. Chemistry can
remain strongly expressed in the dimensional wall state and wall fluxes while producing
similar relative changes in the two normalised transfer coefficients; the small RAF response
results from this similarity between the momentum and thermal transfer responses.

The wider frozen-to-fast-chemistry comparison also illustrates the non-monotonicity
established in \S\ref{sec:nonmono}. Over the full span the dimensional wall heat flux and
shear stress usually recover to same-sign net changes, whereas the intermediate
frozen-to-$\Gamma{=}100$ increments are predominantly of opposite sign. The RAF therefore
depends on the interval over which the chemistry response is measured, and a quoted RAF
sensitivity must specify the corresponding reaction-rate range.

\subsection{Exact decomposition of the RAF response}
\label{sec:raf}

The wall heat flux of a compressible boundary layer can itself be decomposed into generation
and transport contributions \citep{ricco2023decomposition}. The cancellation examined here
admits a similar treatment, expressed directly in terms of the momentum and enthalpy wall-law
variables. Define $u^+(y)=u(y)/u_\tau$ and
$H^+(y)=[h(y)-h_w+ru(y)^2/2]/H_\tau$, where $H_\tau=q_w/(\rho_wu_\tau)$ and $r=Pr_e^{1/3}$.
Their matching-height values are $u_m^+=u_e/u_\tau$ and $H_m^+=\Delta h/H_\tau$. Using
$u_\tau^2=\tau_w/\rho_w$, the Reynolds analogy factor can then be written exactly as
\begin{equation}
\RAF \;=\; \frac{2St}{C_f} \;=\; \frac{q_w u_e}{\tau_w \Delta h} \;=\; \frac{u_m^+}{H_m^+},
\label{eq:rafid}
\end{equation}
Equation~\eqref{eq:rafid} shows that the chemistry sensitivity of the RAF is determined by the
relative displacement of the momentum and enthalpy wall laws. To distinguish these wall-law
shifts from changes associated with the density transformation, define
\begin{equation}
F_u=\int_0^{u_m^+}\sqrt{\frac{\rho}{\rho_w}}\,\mathrm{d}u^+,
\qquad
F_h=\int_0^{H_m^+}\sqrt{\frac{\rho}{\rho_w}}\,\mathrm{d}H^+.
\end{equation}
where the momentum form is the Van Driest density transformation
\citep{vandriest1951compressible}, together with the transformation factors
$\Phi_u=F_u/u_m^+$ and $\Phi_h=F_h/H_m^+$. Taking the logarithmic increment of
\eqref{eq:rafid} gives the exact identity
\begin{equation}
\delta\ln\RAF=\left(\delta\ln F_u-\delta\ln F_h\right)-\left(\delta\ln\Phi_u-\delta\ln\Phi_h\right).
\label{eq:decomp}
\end{equation}
The first difference measures the relative chemistry-induced displacement of the transformed
momentum and enthalpy wall laws; the second measures the differential change introduced by
their density-transformation factors. This decomposition separates two possible sources of RAF
sensitivity without introducing an approximation.

Table~\ref{tab:decomp} shows that the first contribution dominates. Over the full
frozen-to-fast-chemistry span the absolute difference associated with $F_u$ and $F_h$ reaches
$4.290\,\%$, whereas the corresponding $\Phi$ contribution reaches only $0.450\,\%$; the
resulting maximum $|\delta\ln\RAF|$ is $4.308\,\%$. The same ordering holds for the
frozen-to-$\Gamma{=}100$ comparison, for which the maximum contributions are $2.399$, $0.351$
and $2.119\,\%$.
\begin{table}
  \centering
  \begin{tabular}{lccc}
  \hline\hline
  Quantity & median & 95th percentile & max \\
  \hline
  \multicolumn{4}{l}{\textit{frozen} $\to$ \textit{fast chemistry} (full chemical span, $n=712$)} \\
  $\rho_w$          & 0.398 & 5.937 & 14.990 \\
  $\Delta h$        & 0.707 & 8.757 & 29.643 \\
  $F$ group         & 3.019 & 3.952 & 4.290 \\
  $\Phi$ group      & 0.085 & 0.337 & 0.450 \\
  $\delta\ln\RAF$   & 2.965 & 3.927 & \textbf{4.308} \\
  \hline
  \multicolumn{4}{l}{\textit{frozen} $\to$ $\Gamma=100$ (transitional span, $n=712$)} \\
  $\rho_w$          & 11.327 & 21.197 & 25.024 \\
  $\Delta h$        & 13.024 & 20.657 & 23.989 \\
  $F$ group         & 0.453 & 1.693 & 2.399 \\
  $\Phi$ group      & 0.113 & 0.298 & 0.351 \\
  $\delta\ln\RAF$   & 0.417 & 1.465 & \textbf{2.119} \\
  \hline\hline
  \end{tabular}
  \caption{Exact decomposition \eqref{eq:decomp} of the Reynolds-analogy-factor response for
  both pairings on the same $712$ conditions ($281$ cold-edge and $431$ hot-edge). Entries are
  absolute paired logarithmic increments expressed in per cent. Wall-state
  changes of up to about $30\,\%$ reduce to a few per cent in $\RAF$, with the
  non-cancelling response concentrated in the $F$ group rather than the $\Phi$ group.}
  \label{tab:decomp}
\end{table}

The small RAF response therefore does not arise primarily because the compressibility
transformations eliminate the chemistry dependence. The dominant attenuation occurs instead
because the transformed momentum and enthalpy wall laws shift by nearly the same relative
amount, so that their difference $\delta\ln F_u-\delta\ln F_h$ is much smaller than the
underlying changes in wall density, enthalpy difference or individual fluxes.

This interpretation is also evident condition by condition. For the frozen-to-$\Gamma{=}100$
comparison the increments in $F_u$ and $F_h$ cluster close to the one-to-one relation, with a
median absolute separation of $0.307\,\%$ and a ninety-fifth-percentile separation of
$1.552\,\%$. Even when the thermochemical wall state changes substantially, the transformed
momentum and enthalpy responses remain closely aligned.
\begin{figure}[htbp]
  \centering
  \begin{subfigure}{0.49\textwidth}\caption{}%
    \includegraphics[width=\textwidth]{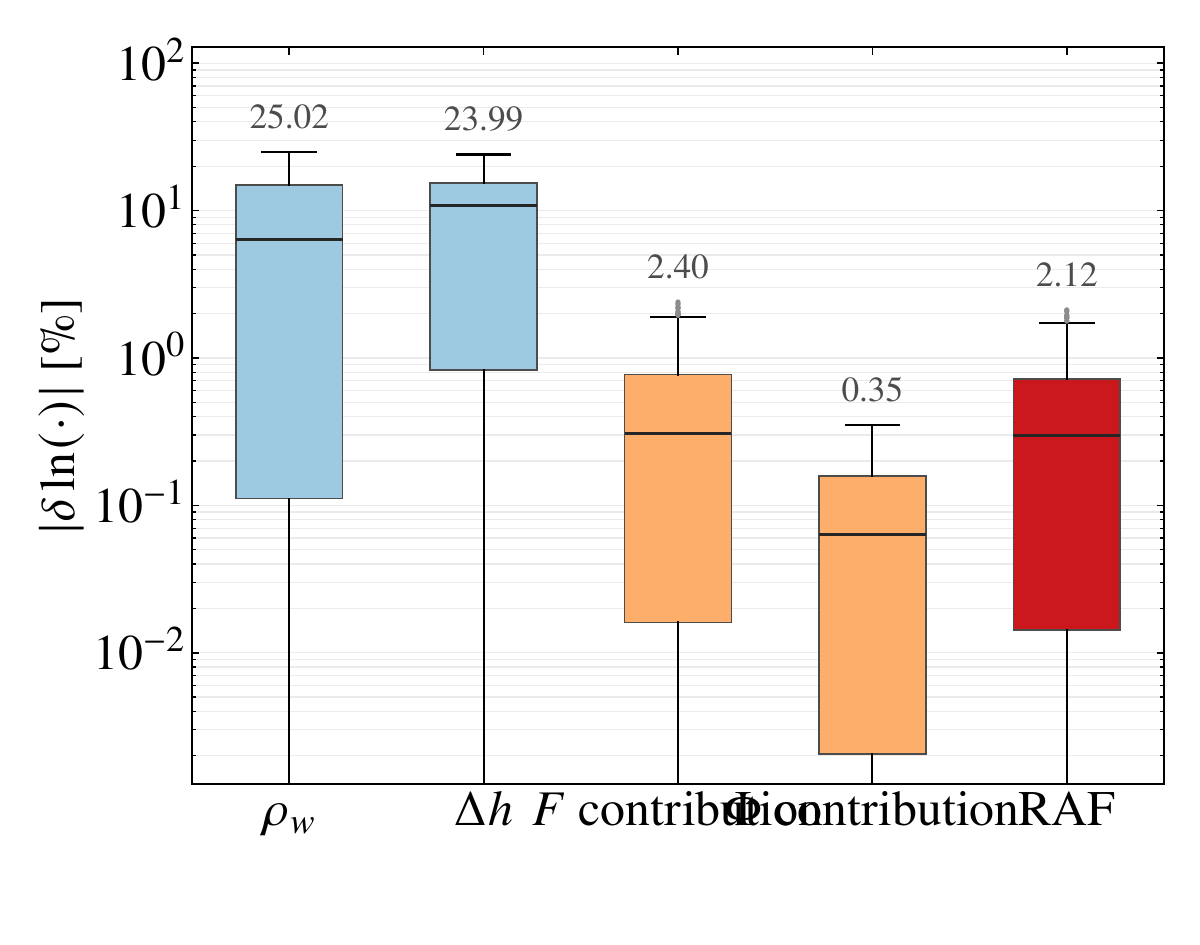}%
    \label{fig:raf-filtering_a}\end{subfigure}\hfill
  \begin{subfigure}{0.49\textwidth}\caption{}%
    \includegraphics[width=\textwidth]{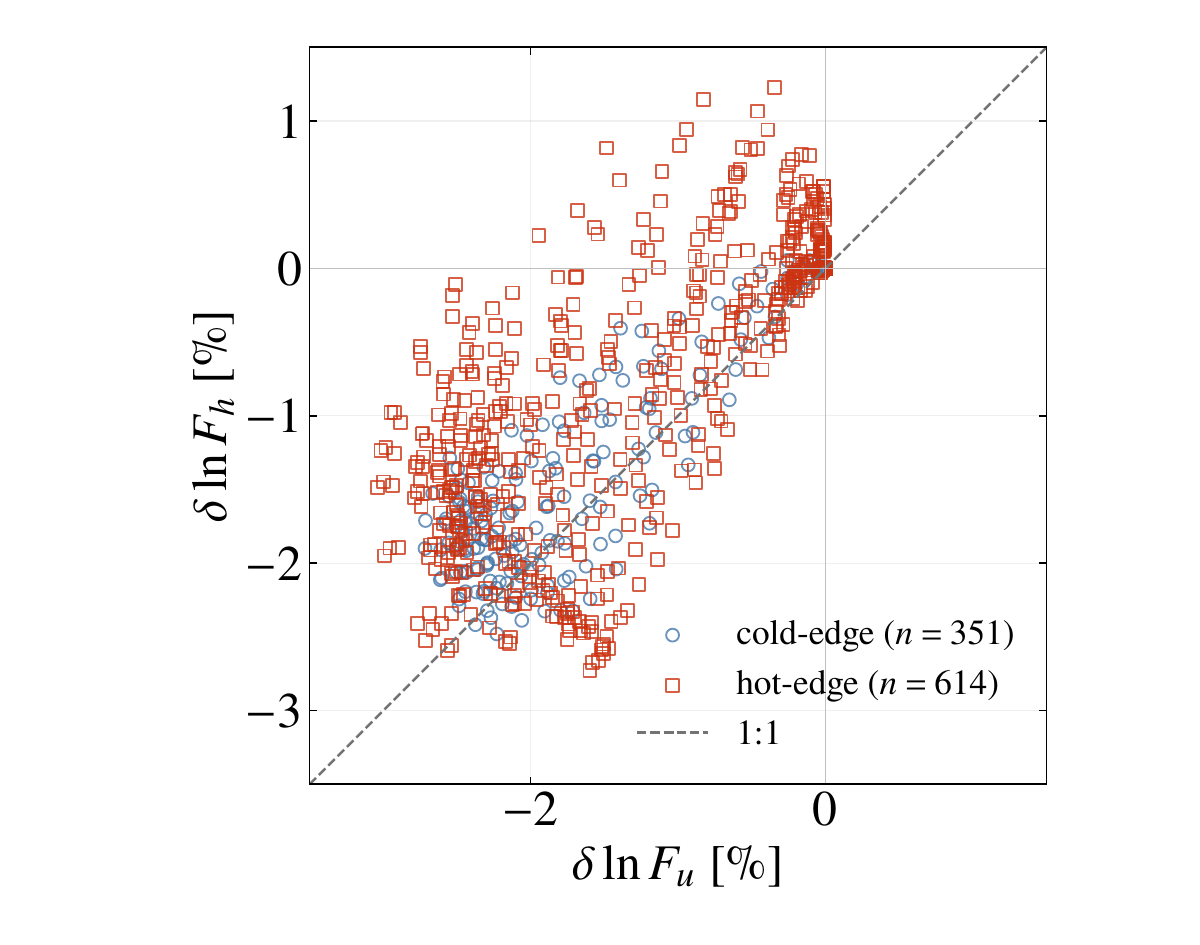}%
    \label{fig:raf-filtering_b}\end{subfigure}
  \caption{Visual summary of the frozen-to-$\Gamma{=}100$ exact decomposition over
  all $965$ available transitional complete-profile pairs ($351$ cold-edge and $614$ hot-edge).
  The set in table~\ref{tab:decomp} is instead the $712$-condition intersection common to both pairing spans.
  (\textit{a}) Distributions of the absolute wall-state increments, the two grouped
  contributions in \eqref{eq:decomp}, and the resulting RAF increment; numbers above the
  boxes are sampled maxima in per cent. (\textit{b}) Condition-by-condition alignment of the
  momentum- and enthalpy-side transformed wall-law increments. Their median and 95th-percentile
  absolute separations from the 1:1 line are $0.307$ and $1.552\,\%$, respectively. The
  wall-state response is therefore large, while the non-cancelling difference between the two
  transformed wall-law shifts remains much smaller.}
  \label{fig:raf-filtering}
\end{figure}

\subsection{Origin and limits of the cancellation}
\label{sec:raf_cancellation}

The alignment identified above is not equivalent to exact Crocco-type linearity. If the
momentum and enthalpy transformations were identical in that stronger sense, $\Phi_h=\Phi_u$
and the second group in \eqref{eq:decomp} would vanish identically. The finite, although
comparatively small, $\Phi$ contribution shows that this condition is not satisfied. The
attenuation instead results mainly from the close but not exact correspondence between the
chemistry-induced shifts in $F_u$ and $F_h$.

The magnitude of the residual depends on how far the reaction rate is varied. On the common
set of conditions used for both comparisons, the maximum sampled RAF response is $2.12\,\%$
between frozen chemistry and $\Gamma=100$, but $4.31\,\%$ between frozen chemistry and the
high-rate limit. Only $17.6\,\%$ of the frozen-to-$\Gamma{=}100$ comparisons exceed a
$1\,\%$ RAF change, whereas $78.4\,\%$ of the full-span comparisons do so. The apparent
robustness of the Reynolds analogy factor is therefore strongest over the transitional
finite-rate range and weakens as the chemistry approaches the high-rate limit.

The same ladders also constrain the fast-chemistry value of $\RAF$. Across the $19$ complete
positive-heat-flux ladders, $|\Delta\ln\RAF|$ from $\Gamma=10^7$ to $10^9$ has a median of
$0.020\,\%$ and a maximum of $0.247\,\%$; the last decade contributes at most $0.063\,\%$.
Sixteen ladders have contracting $\RAF$ increments and give a maximum geometric-tail estimate
of $0.280\,\%$. The other three show less than $0.012\,\%$ drift from $10^7$ to $10^9$ but
do not support tail extrapolation.

Decreasing increments could also result if the grid ceased to resolve the thinning
reaction--diffusion layer. The top two rungs were therefore recomputed on a grid refined by a
factor of two. The maximum coarse--fine differences are $0.024\,\%$ in $q_w$ and
$0.0167\,\%$ in $\RAF$, and neither grows over the last rate decade. The observed contraction
is therefore not caused by progressive loss of wall-normal resolution. Detailed grid ratios
for the ladders are given in the supplementary material.

The principal result is that chemistry can strongly alter wall thermodynamics and the
individual momentum and thermal transfer quantities without producing a comparable change in
their Reynolds-analogy ratio. Under the baseline closure this behaviour is traced to a
near-cancellation between the chemistry-induced shifts of the transformed momentum and
enthalpy wall laws. The cancellation is close rather than exact, leaving a residual response
that depends on the reaction-rate interval. The following sections examine first how the
momentum-side shift can be organised across the sampled conditions and then how turbulent heat
and species transport alter the balance between the two wall-law responses.

\section{Empirical scaling of the momentum-side chemistry response}
\label{sec:scaling}

The exact decomposition in \S\ref{sec:raf} shows that the dominant non-cancelling contribution
to the Reynolds-analogy-factor response is the difference $\delta\ln F_u-\delta\ln F_h$
between the transformed momentum and enthalpy wall-law shifts. The close alignment of these
two quantities under the baseline closure explains the small RAF response, but it does not by
itself identify how either shift varies across the thermodynamic parameter space. The
momentum-side quantity is examined first because it exhibits a particularly simple dependence
on the frozen-state timescale diagnostic and the edge pressure over the
frozen-to-$\Gamma{=}100$ interval,
\begin{equation}
\begin{gathered}
\delta\ln F_u=
\frac{-A}{\left[(p_e/p_{\mathrm{ref}})/\sqrt{\Dafr}\right]^{m}+C},\\
m=1/\ln10,\qquad A=0.1124,\qquad C=3.880,
\end{gathered}
\label{eq:scaling}
\end{equation}
with $p_{\mathrm{ref}}=1$~Pa. The relation describes $687$ paired conditions with
$R^2=0.9821$. When calibrated on the cold-edge conditions and applied to the hot-edge set it
retains $R^2=0.9652$, indicating that the dependence is not simply a consequence of the
particular edge-temperature family used in the fit.

The appearance of $\Dafr^{-1/2}$ is consistent with the reaction--diffusion length scale
discussed in \S\ref{sec:eqlimit}. A balance between diffusion and chemical relaxation gives
$\delta_{rd}\sim\sqrt{Dt_{ch}}$, so a half-power dependence on the chemical timescale provides
a physically motivated coordinate for the momentum response. The additional pressure
dependence in \eqref{eq:scaling} is, however, empirical. The present results do not establish
a universal $Da^{-1/2}$ regime, because $\Dafr$ combines the chemical time with a single local
shear time and does not uniquely represent the full reaction--diffusion balance.

The correlation is also specific to the momentum response over the transitional reaction-rate
interval. Applying the same functional form to the enthalpy-side shift gives substantially
poorer cross-family agreement, with $R^2=0.515$, and applying \eqref{eq:scaling} over the full
frozen-to-fast-chemistry interval reduces the momentum-side agreement to $R^2=0.5445$. The
relation should therefore be read as an empirical description of how the transformed momentum
wall law responds as chemistry changes from frozen to $\Gamma=100$, rather than as a general
wall-law scaling for arbitrary chemical states.

This distinction matters for the RAF response. The momentum-side shift is strongly organised
by the combined pressure--timescale variable in \eqref{eq:scaling}, whereas the enthalpy-side
shift does not obey the same relation. The RAF residual $\delta\ln F_u-\delta\ln F_h$ therefore reflects the part of the
enthalpy-wall-law shift that is not matched by the momentum-wall-law shift. The reaction--diffusion argument, collapse plots, full-span test
and the comparison with alternative functional forms are given in the supplementary material.
\label{sec:scope}

\section{Effects of turbulent heat and species transport}
\label{sec:closure}

The preceding sections show that the finite-rate-chemistry retention criterion is determined
by the wall-heat-flux response, whereas the comparatively small Reynolds-analogy-factor
response results from the close alignment of the transformed momentum and enthalpy wall-law
shifts. Both results depend on the transport closure used in the wall model. To distinguish
the roles of turbulent heat and species transport, the turbulent Prandtl and Schmidt numbers
are varied separately from their baseline values $Pr_t=Sc_t=0.89$.

The two coefficients enter different parts of the governing equations. The turbulent Prandtl
number appears in the effective thermal conductivity
$\lambda_{\mathrm{eff}}=\lambda+\mu_tc_p/Pr_t$ and therefore modifies the temperature and
enthalpy fields directly. The turbulent Schmidt number appears in the species diffusivity
$\rho D_s+\mu_t/Sc_t$ and affects the energy balance indirectly through the species-enthalpy
flux $\sum_sh_sJ_s$. Neither coefficient appears explicitly in the momentum equation.
Varying $Pr_t$ and $Sc_t$ separately therefore isolates the effects of turbulent heat
transport and turbulent species transport on the chemistry response.

\subsection{Sensitivity of the finite-rate-retention criterion}
\label{sec:closure_selection}

The finite-rate-retention boundary responds differently to the two transport coefficients.
Varying $Sc_t$ between $0.60$ and $1.20$ produces comparatively small changes in the
heat-flux-based selection: at most $6.6\,\%$ of the sampled conditions change their four-class
response label of \eqref{eq:zones} relative to the baseline closure, and fewer still change the
binary retention decision itself, while the maximum errors associated with the
limiting chemistry treatments change by no more than approximately $6\,\%$.

This weak sensitivity is also reflected in the temperature-based predictor. Because the frozen
solution contains no species-production response to the rate multiplier, its $T_{\max}$ value
is unchanged by the $Sc_t$ variation considered here, and recalibration of the corresponding
$1\,\%$ criterion moves the fitted threshold only from $3760$ to $3762$~K on the paired set.
No condition changes from frozen adequacy under the baseline closure to finite-rate retention
for $0.60\leq Sc_t\leq1.00$; such a crossing occurs only at the tested extreme $Sc_t=1.20$.

Changing $Pr_t$ displaces the retention boundary substantially more. Because $Pr_t$ directly
modifies turbulent heat transport, it changes the temperature field and the wall-heat-flux
difference used to define the selection criterion, and up to $15.4\,\%$ of the sampled
conditions change their four-class response label over the tested range. The binary retention
decision changes less often, but its crossings are systematic in direction.

The direction of these crossings is systematic about the baseline value. For $Pr_t<0.89$, all
$63$ crossings are from finite-rate retention under the baseline
closure to frozen adequacy under the varied closure: conditions that require finite-rate chemistry under the baseline
closure fall below the tolerance once $Pr_t$ is reduced, so applying the baseline criterion
retains finite-rate chemistry where the varied closure would not demand it. For $Pr_t>0.89$, all $54$ observed crossings occur in the opposite direction:
conditions judged adequately represented by frozen chemistry under the baseline calibration
exceed the finite-rate-retention tolerance after $Pr_t$ is increased. These conditions lie
close to the original selection boundary.

The effect is therefore most important near the transition between frozen adequacy and
finite-rate retention rather than deep within either regime. At the highest Mach numbers
finite-rate chemistry is already required over most of the sampled conditions, leaving
comparatively little scope for the classification to change. Within the present parameter
range the $Pr_t=0.89$ retention criterion should consequently be recalibrated before being
applied at larger turbulent Prandtl numbers.

\subsection{Transport dependence of the RAF cancellation}
\label{sec:closure_raf}

The Reynolds analogy factor exhibits the opposite ordering of sensitivity. The
chemistry-selection boundary is affected more strongly by $Pr_t$, but the small residual left
after momentum--enthalpy cancellation is considerably more sensitive to $Sc_t$.

This follows directly from the exact decomposition of \S\ref{sec:raf}. The momentum-side
quantity $F_u$ is not directly altered by either $Pr_t$ or $Sc_t$ in the present closure,
whereas the enthalpy-side quantity $F_h$ responds to changes in thermal and species transport.
Variations in these coefficients therefore alter the degree to which
$\delta\ln F_h\simeq\delta\ln F_u$, and hence the residual $\delta\ln F_u-\delta\ln F_h$ that
dominates the chemistry-induced RAF response.

The stronger role of species transport is already visible at the profile level. For conditions
available under all tested closures, varying $Sc_t$ changes the $F$-group contribution in the
RAF decomposition by as much as $10.67\,\%$, whereas the corresponding change in the
transformation-factor contribution remains below $1.33\,\%$. The closure dependence therefore
acts primarily through the relative displacement of the transformed momentum and enthalpy wall
laws rather than through the density transformations themselves.
\begin{figure}[htbp]
  \centering
  \begin{subfigure}{0.49\textwidth}\caption{}%
    \includegraphics[width=\textwidth]{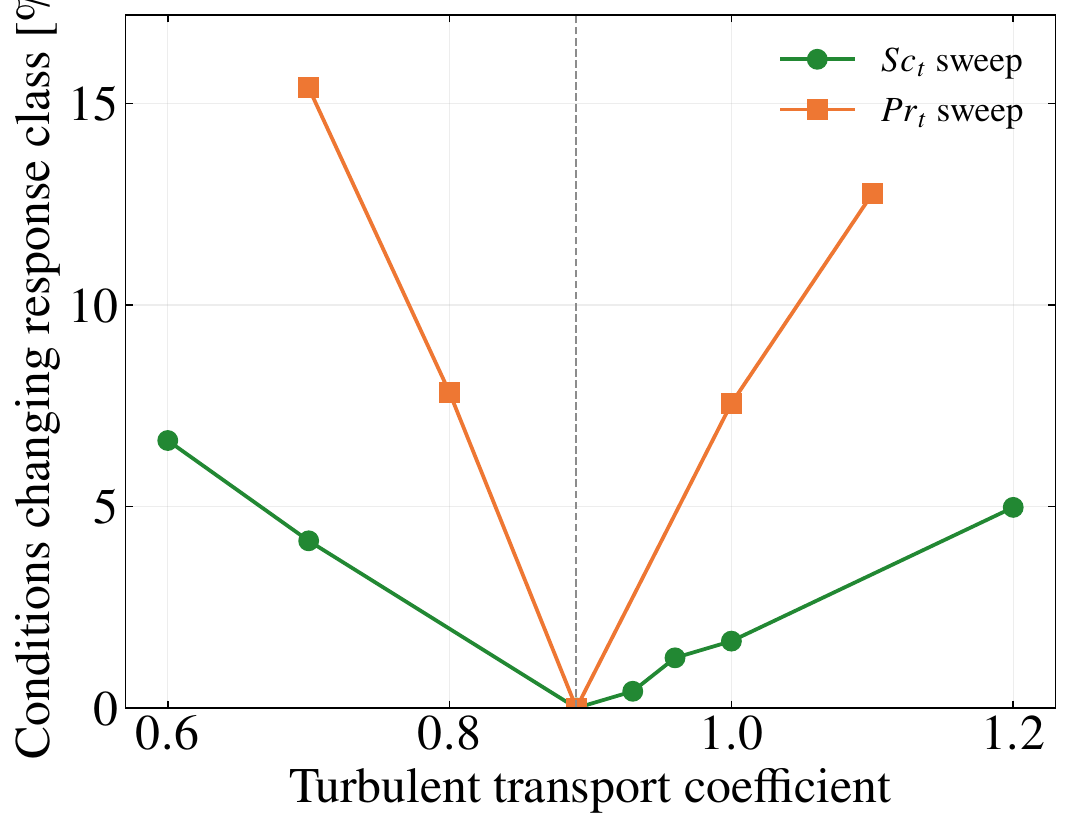}%
    \label{fig:closure-summary_a}\end{subfigure}\hfill
  \begin{subfigure}{0.49\textwidth}\caption{}%
    \includegraphics[width=\textwidth]{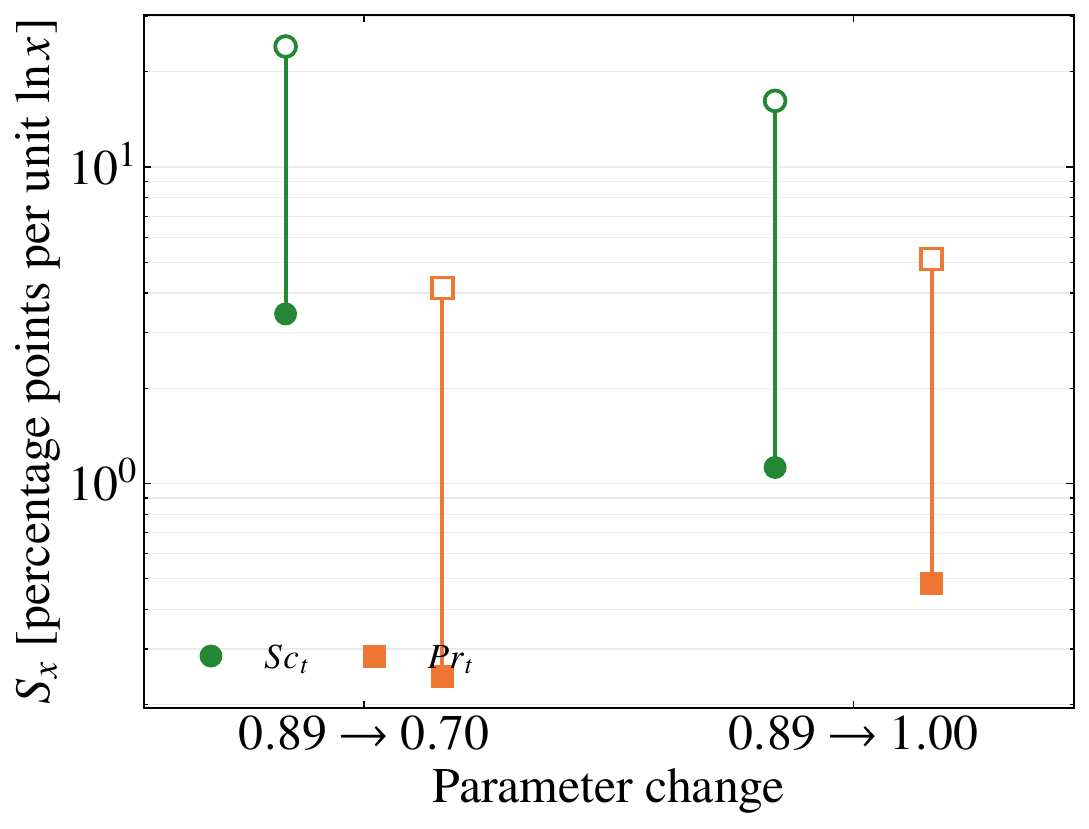}%
    \label{fig:closure-summary_b}\end{subfigure}
  \caption{Complementary effects of turbulent heat and species transport.
  (\textit{a}) Fraction of condition-level response-class assignments changed relative to the
  baseline value $0.89$ within each sweep. The $Sc_t$ values use $241$ paired conditions;
  the $Pr_t$ values use $799$--$807$, so the curves summarise changes within each sweep rather
  than a common-sample comparison. (\textit{b}) Matched finite-difference sensitivity $S_x$
  of the transitional RAF residual on the same $236$ conditions. Filled and open markers are
  the median and 95th percentile, respectively. The selection criterion changes more under the
  tested $Pr_t$ sweep, whereas the RAF residual responds more strongly to matched changes in
  $Sc_t$.}
  \label{fig:closure-summary}
\end{figure}

Across the full $Sc_t$ sweep this change in alignment produces a much larger variation in the
chemistry-induced RAF response than in the heat-flux selection boundary. For the
frozen-to-$\Gamma{=}100$ comparison the largest sampled RAF response has a shallow minimum
near $Sc_t\simeq0.93$ and increases towards both smaller and larger values of $Sc_t$; relative
to the baseline closure the maximum increases by factors of $5.58$ and $3.51$ at the two
sampled extremes. The full frozen-to-fast-chemistry comparison shows the same general increase
away from the baseline region, although its minimum occurs at a different $Sc_t$ because the
full-span difference combines the opposing portions of the non-monotone reaction-rate response
established in \S\ref{sec:nonmono}. The location of these minima should therefore not be
interpreted as a universal optimum turbulent Schmidt number.
\begin{table}
  \centering
  \footnotesize
  \setlength{\tabcolsep}{4pt}%
  \begin{tabularx}{\textwidth}{l c X X}
  \hline\hline
  Sweep & paired $n$ & Map response & Transitional $|\delta\ln\RAF|$ maximum \\
  \hline
  $Sc_t=0.60$--$1.20$ at $Pr_t=0.89$ & $241$
      & $0.4$--$6.6\,\%$ relabelled; one frozen-to-finite-rate crossing at $1.20$
      & $0.82$--$5.58\times$ baseline \\
  $Pr_t=0.70$--$1.10$ at $Sc_t=0.89$ & $799$--$807$
      & up to $15.4\,\%$ relabelled; lower values cross away from finite-rate retention, higher values towards it
      & $0.92$--$2.38\times$ baseline \\
  \hline\hline
  \end{tabularx}
  \caption{Effects of the two turbulent-transport sweeps relative to the baseline closure
  $Pr_t=Sc_t=0.89$. The $Sc_t$ and $Pr_t$ rows use different paired grids and therefore
  summarise within-sweep changes rather than a direct condition-by-condition comparison.
  The $Sc_t$ full-span intersection contains $187$ conditions and shows the same strong
  closure dependence of the sampled $\RAF$ maximum.}
\label{tab:closure-summary}
\end{table}

Changes in the extreme value alone do not describe the full response. Increasing $Sc_t$ from
$0.89$ to $1.00$, for example, raises the largest frozen-to-$\Gamma{=}100$ RAF response by
only a factor of $1.32$, but the number of sampled conditions with a response exceeding
$1\,\%$ increases from $17$ to $83$. Species-transport closure therefore affects not only a
few extreme cases but also the extent of the parameter space over which the RAF becomes
appreciably chemistry-sensitive.

\subsection{Relative roles of heat and species transport}
\label{sec:closure_comparison}

A direct comparison of the two transport channels can be made on the $236$ conditions common
to the baseline closure and to both the $Pr_t$ and $Sc_t$ variations at $0.70$ and $1.00$. Let
$R(x)=|\delta\ln\RAF|$ denote the magnitude of the frozen-to-$\Gamma{=}100$ RAF response,
expressed in percentage points. A finite-difference sensitivity to a transport coefficient $x$
is defined as
\begin{equation}
S_x=\frac{|R(x)-R(0.89)|}{|\ln(x/0.89)|}.
\label{eq:closure_sensitivity}
\end{equation}
For a change from $0.89$ to $0.70$, the median and ninety-fifth-percentile sensitivities are
$3.44$ and $24.1$ for $Sc_t$, compared with $0.246$ and $4.15$ for $Pr_t$. For a change from
$0.89$ to $1.00$ the corresponding values are $1.12$ and $16.2$ for $Sc_t$, compared with
$0.484$ and $5.12$ for $Pr_t$. On the same set of conditions the chemistry-induced RAF
residual is therefore substantially more sensitive to changes in turbulent species transport
than to comparable relative changes in turbulent heat transport.

The two transport coefficients consequently control different aspects of the finite-rate wall
response. The turbulent Prandtl number acts primarily through the thermal field and the wall
heat flux and therefore has the larger influence on the boundary separating frozen adequacy
from finite-rate retention. The turbulent Schmidt number acts directly on species mixing and
the associated enthalpy redistribution, thereby changing the relative momentum and enthalpy
wall-law shifts and exerting the stronger influence on the residual RAF response.

This distinction also explains why the small RAF sensitivity obtained with the baseline
closure is not a closure-independent property of finite-rate chemistry. It results from a
particular near-balance between the momentum and enthalpy responses. Changes in turbulent
species transport can weaken that balance without producing a comparable displacement of the
heat-flux-based chemistry-selection boundary; conversely, changes in turbulent heat transport
can move the selection boundary substantially while leaving the RAF cancellation
comparatively less affected.

The two quantities depend on different parts of the wall-layer response. The
chemistry-selection criterion is controlled mainly by the change in $q_w$, whereas the
RAF is controlled by the difference between the normalised momentum and enthalpy
responses. Their closure sensitivities
should accordingly be assessed separately rather than inferred from one another.

\section{Range of validity}
\label{sec:validity}

The results above apply to the thermochemical and wall-layer model defined in
\S\ref{sec:method}. The principal restrictions arise from the thermochemical temperature
range, the locally parallel wall-model approximation, the turbulent-transport closure and the
definition of the frozen-state timescale diagnostic.

\subsection{Thermochemical and wall-model scope}
\label{sec:validity_scope}

The retained conditions satisfy $T_{\max}<8000$~K, with the hottest cold-edge and hot-edge
cases reaching $7993$ and $7980$~K. Conditions above this limit are excluded because
ionisation may become significant, whereas the present mechanism contains only neutral
species. The resulting chemistry-selection thresholds and reaction-rate sensitivities
therefore apply to the non-ionising, one-temperature thermochemical model considered here
\citep{gupta1990review}.

The temperature cutoff does not imply that all retained conditions are in vibrational
equilibrium. At sufficiently high enthalpy, vibrational nonequilibrium can modify molecular
dissociation and its characteristic timescale before ionisation becomes important. Extending
the present criteria into that regime would therefore require a thermochemical model with
separate internal-energy modes and corresponding finite-rate relaxation processes.

The wall model is steady and locally parallel, with uniform pressure across the modelled layer
and no streamwise pressure-gradient term, and the wall is impermeable and non-catalytic. The
analysis is further restricted to conditions for which $q_w>0$ and $h_{aw}>h_w$, and to states
for which the enthalpy wall-law normalisation remains well defined. These assumptions are
consistent with the controlled wall-layer problem studied here but exclude wall blowing,
catalytic recombination, ablation and strongly streamwise-developing reacting layers.

Species transport is represented by the corrected mixture-averaged approximation introduced in
\S\ref{sec:wallmodel}; multicomponent cross-diffusion and thermal diffusion are not included.
The formulation also contains neither the fluctuating component of the wall fluxes
\citep{yu2022fluctuations} nor the streamwise thermochemical history that can develop through
shocks and compression ramps \citep{guo2025wall}. Both can alter the local chemical state and
the transport timescales and therefore lie outside the present one-dimensional closure.

\subsection{Dependence of the frozen-state criteria on wall-layer scale}
\label{sec:validity_matching}

The primary criteria in \S\ref{sec:criteria} were calibrated using the matching-height
prescription $y_m=1.4\widehat\delta$ at the reference streamwise scale
$x_{\mathrm{ref}}=1$~m. Because $\Dafr$ contains the local shear time, its numerical value is
expected to depend more strongly on the wall-layer scale than a purely thermal indicator such
as $T_{\max}$.

This dependence was examined without refitting the criteria by repeating a set of
thermodynamic conditions at $x_{\mathrm{ref}}=0.5$ and $2$~m. Across these paired
calculations the median $\Dafr$ changes by a factor of $3.37$, whereas the median change in
frozen-state $T_{\max}$ is only $11.2$~K. Despite this difference in the diagnostic variables,
the actual finite-rate-retention decision changes in only one pair: the fixed $\Dafr$
criterion changes its prediction in one pair, while the fixed $T_{\max}$ criterion does not.

The retention boundary itself is therefore less sensitive to the matching-height scale than
the numerical value of $\Dafr$. This does not make the fitted threshold universal. The value
$1.93\times10^3$ remains specific to the present definitions of chemical time, shear time and
sampling location, and to the matching-height prescription, wall model and heat-flux
tolerance.

The distinction is particularly important because the locally parallel equations contain no
streamwise chemical residence time. The success of $\Dafr=t_{shear}/t_{ch}$ therefore supports
the shear time as an effective organising timescale within the present wall-layer model, but
does not establish it as the unique flow timescale for a spatially developing reacting
boundary layer. Residence, diffusion and turbulent-eddy timescales may become equally or more
relevant when streamwise evolution is retained.

\subsection{Dependence on transport closure}
\label{sec:validity_closure}

The chemistry-selection thresholds are also conditional on the turbulent transport closure,
the baseline criteria having been obtained at $Pr_t=Sc_t=0.89$. Section~\ref{sec:closure}
shows that the selection boundary is comparatively insensitive to moderate changes in $Sc_t$
but responds more strongly to $Pr_t$, because the latter directly modifies the thermal field
and the wall heat flux used in the retention criterion.

Within the tested range, applying the baseline calibration at $Pr_t<0.89$ shifts the decision
only in a crossing direction, whereas for $Pr_t>0.89$ some conditions that satisfy the
baseline frozen-adequacy criterion exceed the prescribed finite-rate heat-flux tolerance after
the thermal closure is changed. The baseline threshold should therefore not be assumed
transferable to substantially different turbulent Prandtl numbers without recalibration.

The Reynolds analogy factor has a different closure dependence. Its small chemistry response
under the baseline closure results from the near-alignment of the transformed momentum and
enthalpy wall-law shifts, and \S\ref{sec:closure_raf} shows that this residual is particularly
sensitive to $Sc_t$. The cancellation identified in \S\ref{sec:raf} should therefore be
interpreted as a property of the coupled chemistry and transport closure rather than as a
closure-independent consequence of finite-rate chemistry.

\subsection{Relation to reacting-flow DNS}
\label{sec:validity_dns}

The wall-model response was also compared with a limited set of direct numerical simulation
data covering both non-reacting and reacting conditions. For the reacting comparisons,
absolute profiles and wall quantities differ from the corresponding DNS values by
approximately $1$--$15\,\%$, depending on the quantity and condition.

The available DNS data are not sufficient, however, to validate the quantitative
chemistry-selection thresholds or the detailed momentum--enthalpy cancellation established in
the present parameter study. In the only available paired comparison of a chemistry-induced
skin-friction increment, the DNS and wall model give changes of $-2.11$ and $+0.20\,\%$. The
empirical momentum relation in \S\ref{sec:scaling} and the quantitative RAF cancellation in
\S\ref{sec:raf} should therefore be regarded as results of the present wall-model framework
until they are tested against reacting DNS or wall-modelled LES designed to resolve the
corresponding chemistry-rate variations.

Accordingly, the quantitative selection thresholds and RAF relations reported here
should be interpreted within the present wall-model framework. Their extension to
other thermochemical models or transport closures requires validation or recalibration.

\section{Conclusions}
\label{sec:conclusions}
A controlled variation of all forward and reverse reaction rates has been used to isolate
the effect of chemical timescale in a hypersonic reacting wall model without changing the
equilibrium constants. At a 1\% allowable difference in wall heat flux, two quantities
available from the frozen solution provide effective indicators of this transition:
$Da_{\mathrm{fr}}\ge1.93\times10^{3}$ and $T_{\max}\ge3760~\mathrm{K}$.
Their numerical thresholds are specific to the present wall model, transport closure and
heat-flux tolerance, but the need for finite-rate chemistry is better organised by the local
thermochemical state and timescale competition than by Mach number alone.

The principal physical result is that wall heat transfer does not generally evolve
monotonically between the frozen and infinitely fast reaction-rate limits. As the reaction
rates increase from the frozen state, the wall heat flux can first decrease below both
limiting predictions and recover only at larger rates. The finite-rate solution is therefore
not, in general, an interpolation between the two limiting chemistry descriptions, and the
infinitely fast reaction-rate limit is not necessarily a better approximation than frozen
chemistry. Limiting chemistry solutions should consequently be regarded as distinct
asymptotic states rather than as bounds from which the intermediate finite-rate wall
response can be inferred.

Finite-rate chemistry produces a different signature in the Reynolds analogy factor.
Changes in wall thermodynamic state and in the individual momentum and thermal transfer
quantities can be large, while their ratio remains much less sensitive under the baseline
closure. The exact decomposition developed here shows that this attenuation results
primarily from a near-cancellation between the chemistry-induced shifts of the transformed
momentum and enthalpy wall laws, rather than from an absence of thermochemical influence.
The momentum-side shift is strongly organised over the transitional reaction-rate range by
a pressure--timescale correlation, whereas the enthalpy-side response is more strongly
affected by turbulent transport. Accordingly, $Pr_t$ has the larger effect on the
finite-rate-retention boundary, while $Sc_t$ has the stronger influence on the residual
Reynolds-analogy-factor response. The apparent robustness of the RAF is therefore not
independent of the transport closure.

These conclusions apply to the present non-ionising, one-temperature, steady and locally
parallel wall-layer formulation with the specified turbulent-transport model. Within this
framework, the results show that finite-rate wall response depends on how chemical
relaxation interacts with near-wall transport, while the weaker Reynolds-analogy-factor
response arises from the coupled momentum and enthalpy response.

\begin{acknowledgments}
This work was supported by the China Postdoctoral Science Foundation (Grant No.
2026M794657) and the Postdoctoral Research Funding of Hangzhou International
Innovation Institute of Beihang University (Grant No. 2025BKZ044).

\end{acknowledgments}

\section*{AUTHOR DECLARATIONS}

\subsection*{Conflict of Interest}
The author has no conflicts to disclose.

\subsection*{Author Contributions}
\textbf{Jingchao Zhang}: Conceptualization (lead); Methodology (lead); Software (lead);
Validation (lead); Formal analysis (lead); Investigation (lead); Data curation (lead);
Visualization (lead); Writing -- original draft (lead); Writing -- review \& editing (lead).

\section*{DATA AVAILABILITY}
The processed data that support the findings of this study are available from the
corresponding author upon reasonable request. The wall-model source code is not publicly
available.

\appendix
\makeatletter
\@addtoreset{figure}{section}
\@addtoreset{table}{section}
\providecommand*{\theHsection}{}\renewcommand*{\theHsection}{appendix.\Alph{section}}
\makeatother
\renewcommand{\thefigure}{\Alph{section}.\arabic{figure}}
\renewcommand{\thetable}{\Alph{section}.\arabic{table}}

\FloatBarrier
\bibliography{p4_pof}

\end{document}